\documentclass[amsmath,amssymb,prd,a4paper,twocolumn]{revtex4-1} 

\usepackage{braket}
\usepackage[utf8]{inputenc} %
\usepackage{dcolumn}
\usepackage{bm}
\usepackage{epsfig}
\usepackage{braket}

\usepackage{graphicx}
\usepackage{amsmath}
\usepackage{epstopdf}
\usepackage{amsfonts,amssymb}
\usepackage[utf8]{inputenc}

\usepackage{pstricks}
\usepackage{color}

\usepackage[compat=1.0.0]{tikz-feynman}
\usepackage{simplewick}
\usepackage{color, colortbl}
\definecolor{Gray}{gray}{0.9}
\usepackage{float}
\usepackage{tikz-feynman}
\usepackage{feynmp}
\usepackage{forest}
\usepackage{ulem}
\usepackage{cancel}

\def\beq{\begin{equation}}
\def\eeq{\end{equation}}
\def\beqa{\begin{eqnarray}}
\def\eeqa{\end{eqnarray}}
\def\MeV{\nobreak\,\mbox{MeV}}

\def\gappeq{\mathrel{\rlap {\raise.5ex\hbox{$>$}}
{\lower.5ex\hbox{$\sim$}}}}
\def\lappeq{\mathrel{\rlap{\raise.5ex\hbox{$<$}}
{\lower.5ex\hbox{$\sim$}}}}
\def\Toprel#1\over#2{\mathrel{\mathop{#2}\limits^{#1}}}

\def\enq{\end{equation}}

\begin{document}

\title{The $D^*/D$ ratio in heavy ion collisions}

\author{ L. M. Abreu}
\email{luciano.abreu@ufba.br}
\affiliation{ Instituto de F\'isica, Universidade Federal da Bahia,
Campus Universit\'ario de Ondina, 40170-115, Bahia, Brazil}

\author{F. S. Navarra}
\email{navarra@if.usp.br}
\affiliation{Instituto de F\'{\i}sica, Universidade de S\~{a}o Paulo, \\
Rua do Mat\~ao, 1371, CEP 05508-090,  S\~{a}o Paulo, SP, Brazil}

\author{H. P. L. Vieira}
\email{hilde\underline{ }son@hotmail.com}
\affiliation{ Instituto de F\'isica, Universidade Federal da Bahia,
Campus Universit\'ario de Ondina, 40170-115, Bahia, Brazil}

\begin{abstract}
In this work we study the $D^*$ and $D$ multiplicities and how they change 
during the hadron gas phase of heavy ion collisions. With the help of
an effective Lagrangian formalism, we calculate the production and absorption  
cross sections of the $D^*$ and $D$ mesons in a hadronic medium. We compute
the time evolution of the abundances and the ratio $D^* /D$. They  are
approximately constant in time. Also, assuming a Bjorken type cooling 
and using an empirical relation between the freeze-out temperature and the
central multiplicity density, we estimate $D^* /D$ as a function of
$ dN /d \eta (\eta =0)$, which represents the system size. We find that, while 
the number of $D^*$'s and $D$'s  grows significantly with the system size, their
ratio remains approximately constant. This prediction can be      
compared with future experimental data. Our results suggest that the charm
meson interactions in the hadron gas do not change their multiplicities and
consequently these mesons are close to chemical equilibrium. 
\end{abstract}

\maketitle

\section{Introduction}

It is well accepted that nuclear matter at high temperatures   
or high densities experiences a phase transition to a deconfined
phase of quarks and gluons, the so-called quark-gluon plasma (QGP), in which
the  fundamental degrees of freedom (quarks and gluons) behave as free 
particles  ~\cite{star05,shu,rapp10,pbm}. This deconfined 
medium can be produced in heavy-ion collisions, where a hot and dense QGP
fireball is created. This system  expands and cools down as time evolves. At    
a certain point  quarks, antiquarks and gluons recombine and form a hot hadron 
gas. The properties of this hot and dense system may be accessed through the
study of observables, such as the hadron multiplicities.   

As it has been often pointed out, heavy mesons are of great importance.
They contain heavy quarks, which are produced in the early stages
of the collision. They can carry unique information about the QGP, unlike
light mesons that can be produced in the hadronic medium at later stages.
However,  charm mesons can be created and destroyed in
collisions with the comoving mesons in the hadron gas. For this reason, the
knowledge of the interactions  of $D^{*}$  and $D$ mesons with light mesons
is crucial for the calculation of the final abundances of such particles.
With this motivation, the propagation of heavy-light mesons in a thermal
medium has been intensively studied in different frameworks (we refer the
reader to Ref.~\cite{prino16} for a review).

In the hadronic phase, the different species of hadrons undergo inelastic
reactions, and their  multiplicities can be modified until the  kinetic     
freeze-out, when there are no more interactions and the particles travel to    
reach the detectors. After hadronization and before the kinetic freeze-out, the
hadron gas may be in a state where the production and absorption reactions
occur with an equal rate and hence particle abundances do not change any more.
This is called chemical equilibrium.  In principle the emergence of chemical
equilibrium could be studied numerically. Knowing the multiplicities of the   
particles of all species at the hadronization time and also knowing all their
interaction cross sections,
we could write and solve a set of coupled rate equations. We would then find the
multiplicities as a function of time and we could determine the chemical
freeze-out time (and the corresponding chemical freeze-out temperature), i.e.,
the time from which on all the multiplicites are frozen. Unfortunately, this
calculation can not be done because we do not know all the required cross
sections. However, we have now a reasonable knowledge of some subsets of the
cross sections and we can find partial solutions for the general equilibration
problem.  We can, for example, determine what happens to $K^*$ in the hadron gas
phase, as it was done in \cite{chiara21}. Alternatively, we can postulate
that chemical equilibrium is reached and use statistical mechanics to compute,
in terms of a small number of parameters, the particle multiplicities, as it
is done in the Statistical Hadronization Model (SHM) \cite{shm}.   
The success of the SHM in reproducing the observed multiplicities strongly
suggests that hadron gas formed in heavy ion collisions is indeed in chemical
equilibrium. However, this hypothesis, 
should at some point, be confirmed by microscopic calculations. In the charm
sector, there has been a continuous progress in the study of the interaction
cross sections of charm mesons.  One of the goals of this work is to study the
evolution of the $D^*$ and $D$ populations with a microscopic approach and
check whether these mesons reach  chemical equilibrium or not. 

Recently, the production of  prompt $D^0, D^+ $ and $D^{*+}$  mesons at
midrapidity $(|y|<0.5)$ in Pb–Pb collisions, at the centre-of-mass energy per
nucleon–nucleon pair $\sqrt{s_{NN}} = 5.02 $ TeV has been investigated by the
ALICE collaboration~\cite{ALICE:2021rxa}. Unfortunately, the sistem size
(represented by $ dN / d\eta (\eta = 0)$) dependence  of the $D^* /D$ ratio
was not yet reported, as it was for strange mesons in~\cite{ALICE:2019xyr}. 
This measurement is very interesting since different system sizes represent
different samples of hadron gas with different lifetimes. Longer living systems
are most likely to reach chemical equilibirum, whereas short living systems
are less likely. 

In a previous work, we have investigated the strange sector by evaluating
the ratio between the $K^*$ and $K$ yields $(K^*/K )$ as a function of the
proper time, using the thermal  cross sections of the  interactions
of the $K^{(*)}$ mesons with other light mesons as input in the rate
equations~\cite{chiara21}. The obtained ratio was in very good
agreement with experimental data~\cite{ALICE:2019xyr}. A similar analysis
in the charm sector is going to be presented in what follows. 

The purpose of this  work is to extend the formalism used in 
\cite{chiara21} to evaluate the time evolution of the ratio 
$D^* /D$ during the hadron gas phase of heavy ion collisions. We calculate 
the production and absorption cross sections of the $D^*$ and $D$ mesons 
with the help of an effective Lagrangian formalism, and use them 
in  rate equations to compute the time evolution of the $D^*$ and $D$     
abundances and the ratio $D^* /D$. Finally we estimate $D^* /D$ as a function
of the central multiplicity density ($ dN /d \eta (\eta =0)$). 

The paper is organized as follows. In Section~\ref{CrSec} we describe
the effective formalism and introduce the thermally averaged cross sections 
of the $D^{(*)}$-absorption and production reactions. In Section~\ref{EvolEq} 
we present and analyze the time evolution of the ratio $D^* /D$ and its
relation with the central multiplicity density. Finally,              
Section~\ref{Conclusions} is devoted to the summary of the main points  
and to the concluding remarks. In Appendix \label{Ampl} the explicit  
expressions for the contributions to the amplitudes of the considered  
processes are given.

\section{Interactions of $D$ and $D^*$ with light mesons}
\label{CrSec}
\subsection{Effective Lagrangians and reactions}

\begin{figure}
\begin{tikzpicture}
\begin{feynman}
\vertex (a1) {$D^{*} (p_1)$};
	\vertex[right=1.5cm of a1] (a2);
	\vertex[right=1.cm of a2] (a3) {$\rho   (p_3)$};
	\vertex[right=1.4cm of a3] (a4) {$D (p_1)$};
	\vertex[right=1.5cm of a4] (a5);
	\vertex[right=1.cm of a5] (a6) {$\pi (p_{3})$};
\vertex[below=1.5cm of a1] (c1) {$\bar{D}^{*} (p_2)$};
\vertex[below=1.5cm of a2] (c2);
\vertex[below=1.5cm of a3] (c3) {$\rho (p_4)$};
\vertex[below=1.5cm of a4] (c4) {$\bar{D} (p_2)$};
\vertex[below=1.5cm of a5] (c5);
\vertex[below=1.5cm of a6] (c6) {$\pi (p_4)$};
	\vertex[below=2cm of a2] (d2) {(5.d)};
	\vertex[below=2cm of a5] (d5) {(6.a)};
\diagram* {
(a1) -- (a2), (a2) -- (c3), (c1) -- (c2), (c2) -- (a3), (a2) -- [fermion, edge label'= $\bar{D}$] (c2), (a4) -- (a5), (a5) -- (a6), (c4) -- (c5), (c5) -- (c6), (a5) -- [fermion, edge label'= $\bar{D}^{*}$] (c5)
}; 
\end{feynman}
\end{tikzpicture}

\begin{tikzpicture}
\begin{feynman}
\vertex (a1) {$D (p_1)$};
	\vertex[right=1.5cm of a1] (a2);
	\vertex[right=1.cm of a2] (a3) {$\pi (p_3)$};
	\vertex[right=1.4cm of a3] (a4) {$D (p_1)$};
	\vertex[right=1.5cm of a4] (a5);
	\vertex[right=1.cm of a5] (a6) {$\rho (p_{3})$};
\vertex[below=1.5cm of a1] (c1) {$\bar{D} (p_2)$};
\vertex[below=1.5cm of a2] (c2);
\vertex[below=1.5cm of a3] (c3) {$\pi (p_4)$};
\vertex[below=1.5cm of a4] (c4) {$\bar{D} (p_2)$};
\vertex[below=1.5cm of a5] (c5);
\vertex[below=1.5cm of a6] (c6) {$\pi (p_4)$};
	\vertex[below=2cm of a2] (d2) {(6.b)};
	\vertex[below=2cm of a5] (d5) {(7.a)};
\diagram* {
(a1) -- (a2), (a2) -- (c3), (c1) -- (c2), (c2) -- (a3), (a2) -- [fermion, edge label'= $\bar{D}^{*}$] (c2), (a4) -- (a5), (a5) -- (a6), (c4) -- (c5), (c5) -- (c6), (a5) -- [fermion, edge label'= $\bar{D}$] (c5)
}; 
\end{feynman}
\end{tikzpicture}

\begin{tikzpicture}
\begin{feynman}
\vertex (a1) {$D (p_1)$};
	\vertex[right=1.5cm of a1] (a2);
	\vertex[right=1.cm of a2] (a3) {$\rho (p_3)$};
	\vertex[right=1.4cm of a3] (a4) {$D (p_1)$};
	\vertex[right=1.5cm of a4] (a5);
	\vertex[right=1.cm of a5] (a6) {$\rho (p_{3})$};
\vertex[below=1.5cm of a1] (c1) {$\bar{D} (p_2)$};
\vertex[below=1.5cm of a2] (c2);
\vertex[below=1.5cm of a3] (c3) {$\rho (p_4)$};
\vertex[below=1.5cm of a4] (c4) {$\bar{D} (p_2)$};
\vertex[below=1.5cm of a5] (c5);
\vertex[below=1.5cm of a6] (c6) {$\pi (p_4)$};
	\vertex[below=2cm of a2] (d2) {(7.b)};
	\vertex[below=2cm of a5] (d5) {(7.c)};
\diagram* {
(a1) -- (a2), (a2) -- (c3), (c1) -- (c2), (c2) -- (a3), (a2) -- [fermion, edge label'= $\bar{D}$] (c2), (a4) -- (a5), (a5) -- (a6), (c4) -- (c5), (c5) -- (c6), (a5) -- [fermion, edge label'= $\bar{D}^{*}$] (c5)
}; 
\end{feynman}
\end{tikzpicture}

\begin{tikzpicture}
\begin{feynman}
\vertex (a1) {$D (p_1)$};
	\vertex[right=1.5cm of a1] (a2);
	\vertex[right=1.cm of a2] (a3) {$\rho (p_3)$};
\vertex[below=1.5cm of a1] (c1) {$\bar{D} (p_2)$};
\vertex[below=1.5cm of a2] (c2);
\vertex[below=1.5cm of a3] (c3) {$\rho (p_4)$};
	\vertex[below=2cm of a2] (d2) {(7.d)};
\diagram* {
(a1) -- (a2), (a2) -- (c3), (c1) -- (c2), (c2) -- (a3), (a2) -- [fermion, edge label'= $\bar{D}^{*}$] (c2)
}; 
\end{feynman}
\end{tikzpicture}
\caption{
Diagrams contributing to the processes $D^* \bar{D}^* \to \rho \rho $ [diagram (5d)], $D \bar{D} \to \pi \pi $ [diagrams (6a) and (6b)], and $D \bar{D} \to \rho \rho $ [diagrams (7a)-(7d)], without specification of the charges of the particles. }
\label{DIAG2}
\end{figure}
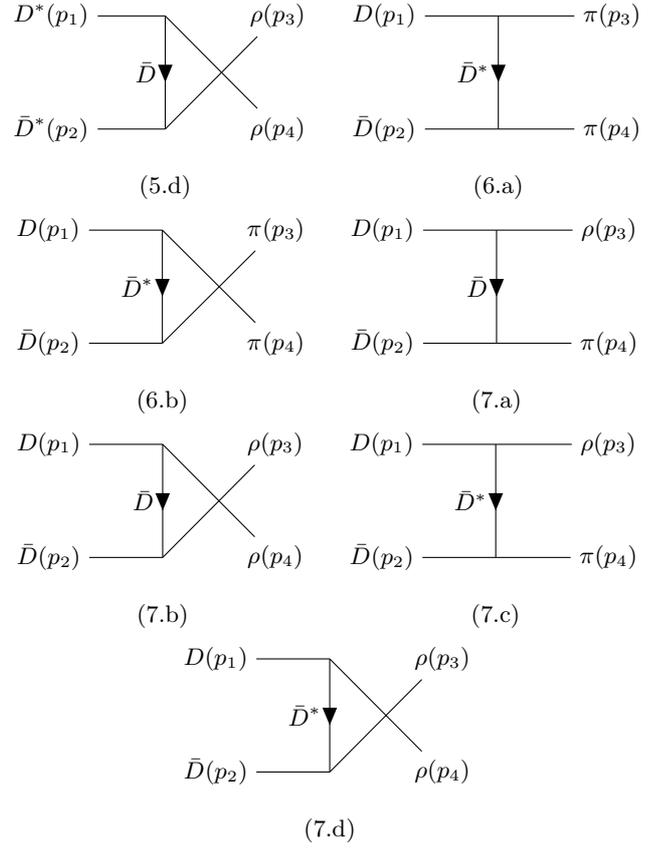

In the present study, the reactions involving the interactions of $D$ and $D^*$ 
mesons with $\pi$ and $\rho$ mesons as well as between them will be analyzed
within an effective field theory approach. In particular, we focus on the  
lowest-order Born contributions to the $D^*$ and $D$ absorption reactions 
shown in Figs.~\ref{DIAG1} and~\ref{DIAG2}, as well as their inverse processes.
To calculate their respective cross sections, we follow                        
Refs.~\cite{suhoung,abreu,chiara21} and employ the effective Lagrangians 
involving $\pi$, $\rho$, $D$  and $D^*$
mesons~\cite{Chen:2007zp,oh,babi05,ChoLee1,torres14,abreu16},
\begin{eqnarray} 
\mathcal{L}_{\pi D D^*} & = & ig_{\pi D D^*} D^{* \mu} \vec{\tau} \cdot (\bar{D} \partial_\mu \vec{\pi} - \partial_{\mu} \bar{D} \vec{\pi}) + h.c., \nonumber \\ 
\mathcal{L}_{\rho DD} & = & ig_{\rho DD} (D \vec{\tau } \partial_{\mu} \bar{D} - \partial_{\mu} D \vec{\tau} \bar{D}) \cdot \vec{\rho}^{\mu}, \nonumber \\
\mathcal{L}_{\rho D^* D^*} &= & i g_{\rho D^* D^*} \left[ (\partial_{\mu} D^{* \nu} \vec{\tau } \bar{D}^{*}_{\nu} - D^{* \nu} \vec{\tau } \partial_{\mu} \bar{D}^{*}_{\nu}) \cdot \vec{\rho}^{\mu} \right. \nonumber \\
&  & + (D^{* \nu} \vec{\tau } \cdot \partial_{\mu} \vec{\rho}_{\nu} - \partial_{\mu} D^{* \nu} \vec{\tau } \cdot \vec{\rho}_{\nu}) \bar{D}^{* \mu}\nonumber \\
  & & + \left. D^{*\mu} (\vec{\tau } \cdot \vec{\rho}^{\nu} \partial_{\mu} \bar{D}^{*}_{\nu} - \vec{\tau } \cdot \partial_{\mu} \vec{\rho}^{\nu} \bar{D}^{*}_{\nu})\right] ,\nonumber\\ 
\mathcal{L}_{\pi D^{*} D^{*}} &=& -g_{\pi D^* D^*} \epsilon^{\mu \nu \alpha \beta} \partial_{\mu} D^{*}_{\nu} \pi \partial_{\alpha} \bar{D}^{*}_{\beta},\nonumber\\ 
\mathcal{L}_{\rho DD^*} &=& -g_{\rho DD^*} \epsilon^{\mu \nu \alpha \beta} (D \partial_{\mu} \rho_{\nu} \partial_{\alpha} \bar{D}^{*}_{\beta} + \partial_{\mu} D^{*}_{\nu} \partial_{\alpha} \rho_{\beta} \bar{D})
	\nonumber \\
\label{EffLagf}
\end{eqnarray}
where $\vec{\tau}$ are the Pauli matrices in the isospin space; $\vec{\pi}$   
denotes the pion isospin triplet; and $D^{(\ast)} = (D^{(\ast) 0}, D^{(\ast)+})$
represents the isospin doublets for the pseudoscalar (vector) $D^{(\ast) }$  
mesons. The coupling constants $g_{\pi DD^*}$, $g_{\rho DD}$,
$g_{\rho D^* D^*}$, $g_{\pi D^* D^*}$ and $g_{\rho D D^*}$ will have their
values given in the next section.

  
\begin{figure}
\centering
\begin{tikzpicture}
\begin{feynman}
\vertex (a1) {$D^{*}  (p_1)$};
	\vertex[right=1.5cm of a1] (a2);
	\vertex[right=1.cm of a2] (a3) {$\rho (p_3)$};
	\vertex[right=1.4cm of a3] (a4) {$D^{*} (p_1)$};
	\vertex[right=1.5cm of a4] (a5);
	\vertex[right=1.cm of a5] (a6) {$\rho (p_{3})$};
\vertex[below=1.5cm of a1] (c1) {$\pi (p_2)$};
\vertex[below=1.5cm of a2] (c2);
\vertex[below=1.5cm of a3] (c3) {$D (p_4)$};
\vertex[below=1.5cm of a4] (c4) {$\pi (p_2)$};
\vertex[below=1.5cm of a5] (c5);
\vertex[below=1.5cm of a6] (c6) {$D (p_4)$};
	\vertex[below=0.75cm of a4] (b4);
\vertex[right=1cm of b4] (b5);
\vertex[right=1cm of b5] (b6);
	\vertex[below=2cm of a2] (d2) {(1.a)};
	\vertex[below=2cm of a5] (d5) {(1.b)};
\diagram* {
(a1) -- (a2), (a2) -- (a3), (c1) -- (c2), (c2) -- (c3), (a2) -- [fermion, edge label'= $D^{*}$] (c2), (a4) -- (b5), (b6) -- (a6), (c4) -- (b5), (b6) -- (c6), (b5) -- [fermion, edge label'= $D$] (b6)
}; 
\end{feynman}
\end{tikzpicture}

\begin{tikzpicture}
\begin{feynman}
\vertex (a1) {$D^{*} (p_1)$};
	\vertex[right=1.5cm of a1] (a2);
	\vertex[right=1.cm of a2] (a3) {$\rho  (p_3)$};
	\vertex[right=1.4cm of a3] (a4) {$D^{*} (p_1)$};
	\vertex[right=1.5cm of a4] (a5);
	\vertex[right=1.cm of a5] (a6) {$\pi  (p_{3})$};
\vertex[below=1.5cm of a1] (c1) {$\pi (p_2)$};
\vertex[below=1.5cm of a2] (c2);
\vertex[below=1.5cm of a3] (c3) {$D (p_4)$};
\vertex[below=1.5cm of a4] (c4) {$\rho (p_2)$};
\vertex[below=1.5cm of a5] (c5);
\vertex[below=1.5cm of a6] (c6) {$D (p_4)$};
	\vertex[below=0.75cm of a1] (b1);
\vertex[right=1cm of b1] (b2);
\vertex[right=1cm of b2] (b3);
	\vertex[below=2cm of a2] (d2) {(1.c)};
	\vertex[below=2cm of a5] (d5) {(2.a)};
\diagram* {
(a1) -- (b2), (b3) -- (a3), (c1) -- (b2), (b3) -- (c3), (b2) -- [fermion, edge label'= $\bar{D}^{*}$] (b3), (a4) -- (a5), (a5) -- (a6), (c4) -- (c5), (c5) -- (c6), (a5) -- [fermion, edge label'= $\bar{D}$] (c5)
}; 
\end{feynman}
\end{tikzpicture}

\begin{tikzpicture}
\begin{feynman}
\vertex (a1) {$D^{*} (p_1)$};
	\vertex[right=1.5cm of a1] (a2);
	\vertex[right=1.cm of a2] (a3) {$\pi (p_3)$};
	\vertex[right=1.4cm of a3] (a4) {$D^{*} (p_1)$};
	\vertex[right=1.5cm of a4] (a5);
	\vertex[right=1.cm of a5] (a6) {$\pi  (p_{3})$};
\vertex[below=1.5cm of a1] (c1) {$\rho (p_2)$};
\vertex[below=1.5cm of a2] (c2);
\vertex[below=1.5cm of a3] (c3) {$D (p_4)$};
\vertex[below=1.5cm of a4] (c4) {$\rho (p_2)$};
\vertex[below=1.5cm of a5] (c5);
\vertex[below=1.5cm of a6] (c6) {$D (p_4)$};
	\vertex[below=0.75cm of a1] (b1);
\vertex[right=1cm of b1] (b2);
\vertex[right=1cm of b2] (b3);
	\vertex[below=0.75cm of a4] (b4);
\vertex[right=1cm of b4] (b5);
\vertex[right=1cm of b5] (b6);
	\vertex[below=2cm of a2] (d2) {(2.b)};
	\vertex[below=2cm of a5] (d5) {(2.c)};
\diagram* {
(a1) -- (b2), (b3) -- (a3), (c1) -- (b2), (b3) -- (c3), (b2) -- [fermion, edge label'= $\bar{D}^{*}$] (b3), (a4) -- (a5), (a5) -- (a6), (c4) -- (c5), (c5) -- (c6), (a5) -- [fermion, edge label'= $\bar{D}^{*}$] (c5)
}; 
\end{feynman}
\end{tikzpicture}

\begin{tikzpicture}
\begin{feynman}
\vertex (a1) {$D^{*} (p_1)$};
	\vertex[right=1.5cm of a1] (a2);
	\vertex[right=1.cm of a2] (a3) {$\rho (p_3)$};
	\vertex[right=1.4cm of a3] (a4) {$D^{*} (p_1)$};
	\vertex[right=1.5cm of a4] (a5);
	\vertex[right=1.cm of a5] (a6) {$\rho  (p_{3})$};
\vertex[below=1.5cm of a1] (c1) {$\bar{D} (p_2)$};
\vertex[below=1.5cm of a2] (c2);
\vertex[below=1.5cm of a3] (c3) {$\pi (p_4)$};
\vertex[below=1.5cm of a4] (c4) {$\bar{D} (p_2)$};
\vertex[below=1.5cm of a5] (c5);
\vertex[below=1.5cm of a6] (c6) {$\pi (p_4)$};
	\vertex[below=2cm of a2] (d2) {(3.a)};
	\vertex[below=2cm of a5] (d5) {(3.b)};
\diagram* {
(a1) -- (a2), (a2) -- (a3), (c1) -- (c2), (c2) -- (c3), (a2) -- [fermion, edge label'= $\bar{D}^{*}$] (c2), (a4) -- (a5), (a5) -- (c6), (c4) -- (c5), (c5) -- (a6), (a5) -- [fermion, edge label'= $\bar{D}$] (c5)
}; 
\end{feynman}
\end{tikzpicture}

\begin{tikzpicture}
\begin{feynman}
\vertex (a1) {$D^{*} (p_1)$};
	\vertex[right=1.5cm of a1] (a2);
	\vertex[right=1.cm of a2] (a3) {$\rho (p_3)$};
	\vertex[right=1.4cm of a3] (a4) {$D^{*} (p_1)$};
	\vertex[right=1.5cm of a4] (a5);
	\vertex[right=1.cm of a5] (a6) {$\pi  (p_{3})$};
\vertex[below=1.5cm of a1] (c1) {$\bar{D} (p_2)$};
\vertex[below=1.5cm of a2] (c2);
\vertex[below=1.5cm of a3] (c3) {$\pi (p_4)$};
\vertex[below=1.5cm of a4] (c4) {$\bar{D}^{*} (p_2)$};
\vertex[below=1.5cm of a5] (c5);
\vertex[below=1.5cm of a6] (c6) {$\pi (p_4)$};
	\vertex[below=2cm of a2] (d2) {(3.c)};
	\vertex[below=2cm of a5] (d5) {(4.a)};
\diagram* {
(a1) -- (a2), (a2) -- (c3), (c1) -- (c2), (c2) -- (a3), (a2) -- [fermion, edge label'= $\bar{D}^{*}$] (c2), (a4) -- (a5), (a5) -- (a6), (c4) -- (c5), (c5) -- (c6), (a5) -- [fermion, edge label'= $\bar{D}$] (c5)
}; 
\end{feynman}
\end{tikzpicture}

\begin{tikzpicture}
\begin{feynman}
\vertex (a1) {$D^{*} (p_1)$};
	\vertex[right=1.5cm of a1] (a2);
	\vertex[right=1.cm of a2] (a3) {$\pi  (p_3)$};
	\vertex[right=1.4cm of a3] (a4) {$D^{*} (p_1)$};
	\vertex[right=1.5cm of a4] (a5);
	\vertex[right=1.cm of a5] (a6) {$\pi  (p_{3})$};
\vertex[below=1.5cm of a1] (c1) {$\bar{D}^{*} (p_2)$};
\vertex[below=1.5cm of a2] (c2);
\vertex[below=1.5cm of a3] (c3) {$\pi (p_4)$};
\vertex[below=1.5cm of a4] (c4) {$\bar{D}^{*} (p_2)$};
\vertex[below=1.5cm of a5] (c5);
\vertex[below=1.5cm of a6] (c6) {$\pi (p_4)$};
	\vertex[below=2cm of a2] (d2) {(4.b)};
	\vertex[below=2cm of a5] (d5) {(4.c)};
\diagram* {
(a1) -- (a2), (a2) -- (c3), (c1) -- (c2), (c2) -- (a3), (a2) -- [fermion, edge label'= $\bar{D}$] (c2), (a4) -- (a5), (a5) -- (a6), (c4) -- (c5), (c5) -- (c6), (a5) -- [fermion, edge label'= $\bar{D}^{*}$] (c5)
}; 
\end{feynman}
\end{tikzpicture}

\begin{tikzpicture}
\begin{feynman}
\vertex (a1) {$D^{*} (p_1)$};
	\vertex[right=1.5cm of a1] (a2);
	\vertex[right=1.cm of a2] (a3) {$\pi  (p_3)$};
	\vertex[right=1.4cm of a3] (a4) {$D^{*} (p_1)$};
	\vertex[right=1.5cm of a4] (a5);
	\vertex[right=1.cm of a5] (a6) {$\rho (p_{3})$};
\vertex[below=1.5cm of a1] (c1) {$\bar{D}^{*} (p_2)$};
\vertex[below=1.5cm of a2] (c2);
\vertex[below=1.5cm of a3] (c3) {$\pi (p_4)$};
\vertex[below=1.5cm of a4] (c4) {$\bar{D}^{*} (p_2)$};
\vertex[below=1.5cm of a5] (c5);
\vertex[below=1.5cm of a6] (c6) {$\rho (p_4)$};
	\vertex[below=2cm of a2] (d2) {(4.d)};
	\vertex[below=2cm of a5] (d5) {(5.a)};
\diagram* {
(a1) -- (a2), (a2) -- (c3), (c1) -- (c2), (c2) -- (a3), (a2) -- [fermion, edge label'= $\bar{D}$] (c2), (a4) -- (a5), (a5) -- (a6), (c4) -- (c5), (c5) -- (c6), (a5) -- [fermion, edge label'= $\bar{D}^{*}$] (c5)
}; 
\end{feynman}
\end{tikzpicture}

\begin{tikzpicture}
\begin{feynman}
\vertex (a1) {$D^{*} (p_1)$};
	\vertex[right=1.5cm of a1] (a2);
	\vertex[right=1.cm of a2] (a3) {$\rho   (p_3)$};
	\vertex[right=1.4cm of a3] (a4) {$D^{*} (p_1)$};
	\vertex[right=1.5cm of a4] (a5);
	\vertex[right=1.cm of a5] (a6) {$\rho (p_{3})$};
\vertex[below=1.5cm of a1] (c1) {$\bar{D}^{*} (p_2)$};
\vertex[below=1.5cm of a2] (c2);
\vertex[below=1.5cm of a3] (c3) {$\rho (p_4)$};
\vertex[below=1.5cm of a4] (c4) {$\bar{D}^{*} (p_2)$};
\vertex[below=1.5cm of a5] (c5);
\vertex[below=1.5cm of a6] (c6) {$\rho (p_4)$};
	\vertex[below=2cm of a2] (d2) {(5.b)};
	\vertex[below=2cm of a5] (d5) {(5.c)};
\diagram* {
(a1) -- (a2), (a2) -- (c3), (c1) -- (c2), (c2) -- (a3), (a2) -- [fermion, edge label'= $\bar{D}^{*}$] (c2), (a4) -- (a5), (a5) -- (a6), (c4) -- (c5), (c5) -- (c6), (a5) -- [fermion, edge label'= $\bar{D}$] (c5)
}; 
\end{feynman}
\end{tikzpicture}
\caption{
Diagrams contributing to the processes $D^* \pi \to \rho D  $ [diagrams (1a)-(1c)], $D^* \rho \to \pi D  $ [diagrams (2a)-(2c)], $D^* \bar{D} \to \rho \pi $ [diagrams (3a)-(3c)], $D^* \bar{D}^* \to \pi \pi $ [diagrams (4a)-(4d)], $D^* \bar{D}^* \to \rho \rho $ [diagrams (5a)-(5c)]. }
\label{DIAG1}
\end{figure}
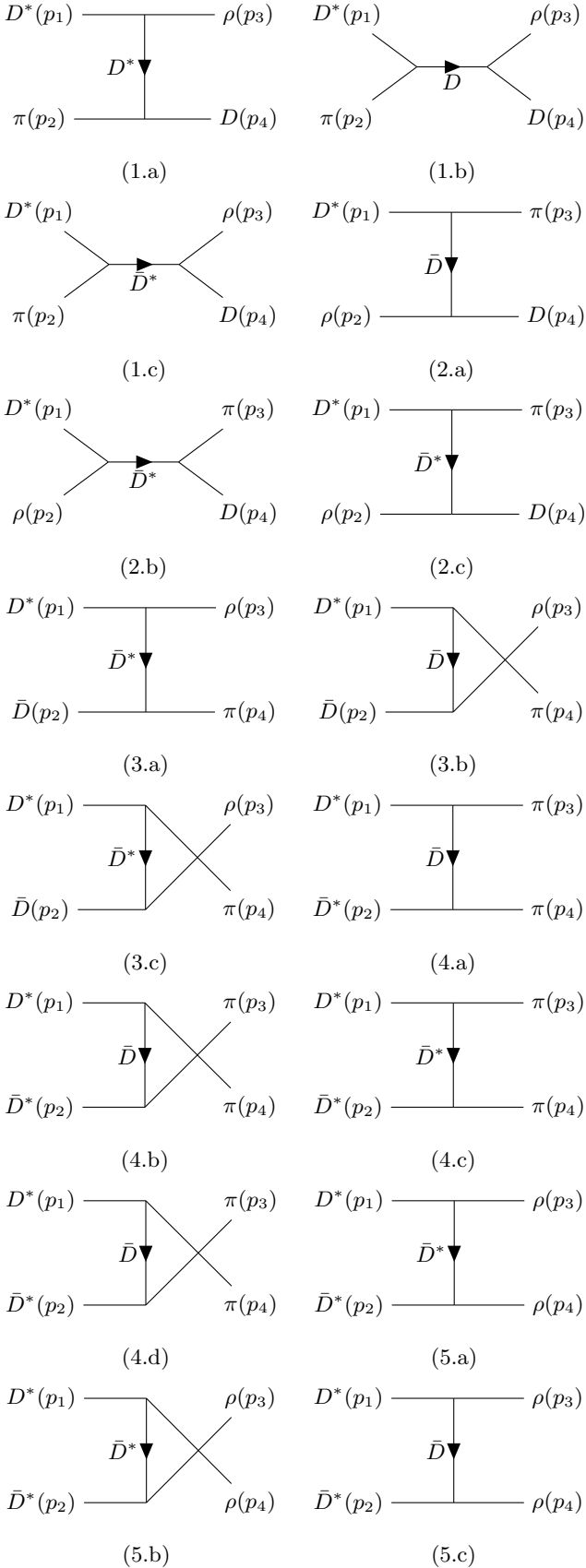

With the effective Lagrangians introduced above, the amplitudes
of the $D^*$ and $D$ absorption processes shown in Figs.~\ref{DIAG1}
and~\ref{DIAG2} can be calculated and are given by 
\begin{eqnarray}
  	\mathcal{M}_{D^* \pi \to \rho D } & = & \mathcal{M}_{(1a)} + \mathcal{M}_{(1b)} + \mathcal{M}_{(1c)},  \nonumber \\
  	\mathcal{M}_{D^* \rho \to \pi D  } & = & \mathcal{M}_{(2a)} + \mathcal{M}_{(2b)} + \mathcal{M}_{(2c)} ,  \nonumber \\
  	\mathcal{M}_{D^* \bar{D} \to \rho \pi } & = & \mathcal{M}_{(3a)} + \mathcal{M}_{(3b)} + \mathcal{M}_{(3c)}, \nonumber \\ 
  	\mathcal{M}_{D^* \bar{D}^* \to \pi \pi } & = & \mathcal{M}_{(4a)} + \mathcal{M}_{(4b)} + \mathcal{M}_{(4c)} + \mathcal{M}_{(4d)},  \nonumber \\
  	\mathcal{M}_{D^* \bar{D}^* \to \rho \rho } & = & \mathcal{M}_{(5a)} + \mathcal{M}_{(5b)} + \mathcal{M}_{(5c)} + \mathcal{M}_{(5d)},  \nonumber \\
  	\mathcal{M}_{D \bar{D} \to \pi \pi } & = & \mathcal{M}_{(6a)} + \mathcal{M}_{(6b)} ,  \nonumber \\
  	\mathcal{M}_{D \bar{D} \to \rho \rho } & = & \mathcal{M}_{(7a)} + \mathcal{M}_{(7b)} + \mathcal{M}_{(7c)} + \mathcal{M}_{(7d)} , 
\label{Amplitudes}
\end{eqnarray} 
where the expressions for each contribution $\mathcal{M}_{(p)}$ are       
explicitly summarized in Appendix~\ref{Ampl}. We mention that, in contrast to 
the works involving the strange mesons~\cite{suhoung,abreu,chiara21},  
here we do not consider the decay width $\Gamma_{D^*}$ in the propagators 
of the intermediate vector charmed mesons, since it is very small and does 
not change  our results significantly.

\subsection{Cross sections}

The isospin-spin-averaged cross section in the center of
mass (CM) frame for the processes in Eq. (\ref{Amplitudes}) is given by: 
\begin{eqnarray}
  \sigma_r ^{\left(\varphi \right)}(s) 
= \frac{1}{64 \pi^2 s }  \frac{|\vec{p}_{f}|}{|\vec{p}_i|}  \int d \Omega 
\overline{\sum_{S, I}} 
|\mathcal{M}_r  ^{\left(\varphi \right)} (s,\theta)|^2 ,
\label{eq:CrossSection}
\end{eqnarray}
where $r $ denominates the reactions according to Eq.~(\ref{Amplitudes}); 
$\sqrt{s}$ is the CM energy;  $|\vec{p}_{i}|$ and $|\vec{p}_{f}|$ denote 
the three-momenta of initial and final particles in the CM frame,            
respectively; the symbol $\overline{\sum_{S,I}}$ stands for the sum over the
spins and isospins of the particles in the initial and 
final state, weighted by the isospin and spin degeneracy factors of the two
particles forming the initial state for the reaction $r$, i.e.
\begin{eqnarray}
\overline{\sum_{S,I}}|\mathcal{M}_r|^2 & \to & 
\frac{1}{g_{1i,r}}
 \frac{1}{g_{2i,r}} \sum_{S,I}|\mathcal{M}_r|^2, 
\label{eq:DegeneracyFactors}
\end{eqnarray}
where $g_{1i,r}=(2I_{1i,r}+1)(2I_{2i,r}+1)$ and
$g_{2i,r}= (2S_{1i,r}+1)(2S_{2i,r}+1)$ are the degeneracy factors  of the
particles in the initial state. The cross sections for the inverse processes
of those shown in Figs.~\ref{DIAG1} and~\ref{DIAG2} can be calculated through
the use of the detailed balance relation.  

In the evaluation of the cross sections, to prevent   
the artificial increase of the amplitudes with the energy and take into     
account the finite size of the hadrons, we introduce form factors in each
vertex in the reactions. Fortunately, all the form factors for
the vertices $D \pi D^* $, $D \rho D$, $D^* \rho D^*$, $D^* \pi D^*$ and
$D \rho D^*$ have already been calculated with the method of QCD sum rules  
in previous works \cite{bcnn11,rdsds,rdsd,vieira22}.  They were parametrized
with the following forms \cite{bcnn11} :
\beq
I) \,\,\,\, g_{M_1 M_2 M_3} = \frac{A}{Q^2 + B}
\label{FI}
\enq
and
\beq
II) \,\,\,\, g_{M_1 M_2 M_3} = A \, e^{-(Q^2/B)} 
\label{FII}
\enq
where $M_1$ is the off-shell meson in the vertex and $Q^2$ is its euclidean
four momentum squared. The parameters $A$ and $B$ are given in
Table \ref{FFparam}. 

\begin{table}[h] 
        \begin{center} 
                \begin{tabular}{cccc} 
                        \hline 
                        \hline 
                        $M_1$ $M_2$ $M_3$ & Form & A    & B     \\ 
                        \hline 
                        $D \pi D^* $      & I    & 126  & 11.9  \\ 
                        \hline 
                        $D \rho D$        & I    & 37.5 & 12.1  \\ 
                        \hline  
                        $D^* \rho D^*$    & II   & 4.9  & 13.3  \\ 
                        \hline  
                        $D^* \pi D^*$     & II   & 4.8  & 6.8   \\ 
                        \hline  
                        $D \rho D^*$      & I    & 234  & 44    \\ 
                        \hline  
                        \hline 
                \end{tabular} 
\caption{Parameters for the form factors in the $M_1$ $M_2$ $M_3$
vertex~\cite{bcnn11}. The meson $M_1$ is off-shell. }
                \label{FFparam} 
        \end{center} 
\end{table}

The final element in this set of reactions is the  formation
of the $ D^* $ meson from the pion and $D$ meson. Adopting a 
similar approach as in Refs.~\cite{suhoung,chiara21,Abreu:2020ony}, the  
scattering cross section for the process $D \pi \to D^*$ is given by the
spin-averaged relativistic Breit-Wigner cross section, 
\begin{eqnarray} 
  \sigma _{D  \pi \to D^*} = \frac{g_{D^*}}{g_{\bar{D} } g_{\pi}}
  \frac{4 \pi}{p_{cm} ^2 }
  \frac{s \Gamma_{D^* \to D  \pi} ^2}{\left( s - m_{D^* }^2 \right)^2
    + s  \Gamma_{D^* \to D \pi} ^2 }, 
  \label{CrSecFormX}
 \end{eqnarray}
where $g_{D^*}, g_{\bar{D} } $ and $g_{\pi}$ are the degeneracy of $D^*$, 
$\bar{D}$ and $\pi$ mesons, respectively; $p_{cm}$ is the momentum in CM  
frame; $\Gamma_{D^* \to D \pi}$ is the total decay width for the reaction 
$D^* \to D \pi$, which is supposed to be effectively $\sqrt{s}$-dependent
via the formula
\begin{eqnarray} 
  \Gamma_{D^* \to D \pi} (\sqrt{s}) = \frac{g_{D^* \to D \pi}^2 }{2 \pi s }
  p_{cm}^3(\sqrt{s}), 
    \label{decayDSTAR}
\end{eqnarray}
with the value of constant $g_{D^* \to D \pi}$ being determined from the    
experimental value of $ \Gamma_{D^* \to D \pi} (\sqrt{s}) $. 

We have employed in the computations of the present work the isospin-averaged 
masses:  $m_{\pi} = 137.3 $ MeV, $m_{\rho} = 775.2 $ MeV, $m_D = 1867.2 $ MeV, 
$m_{D^{\ast}} = 2008.6$ MeV; and for the decay width:
$\Gamma _{D^* \to D \pi} = 69.2$ keV.

\begin{figure}[th]
\centering
\includegraphics[width=1.0\linewidth]{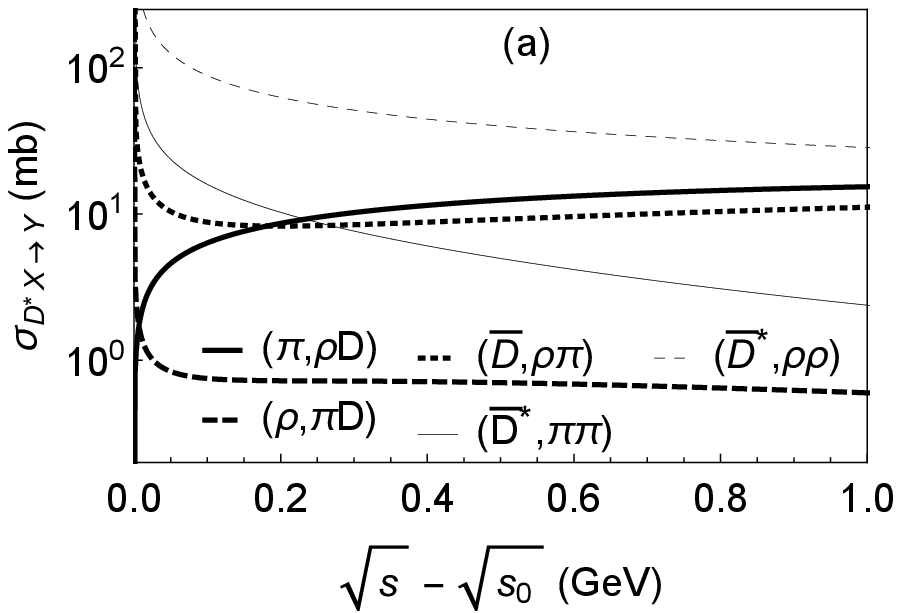}
\\
\vskip0.8cm 
\includegraphics[width=1.0\linewidth]{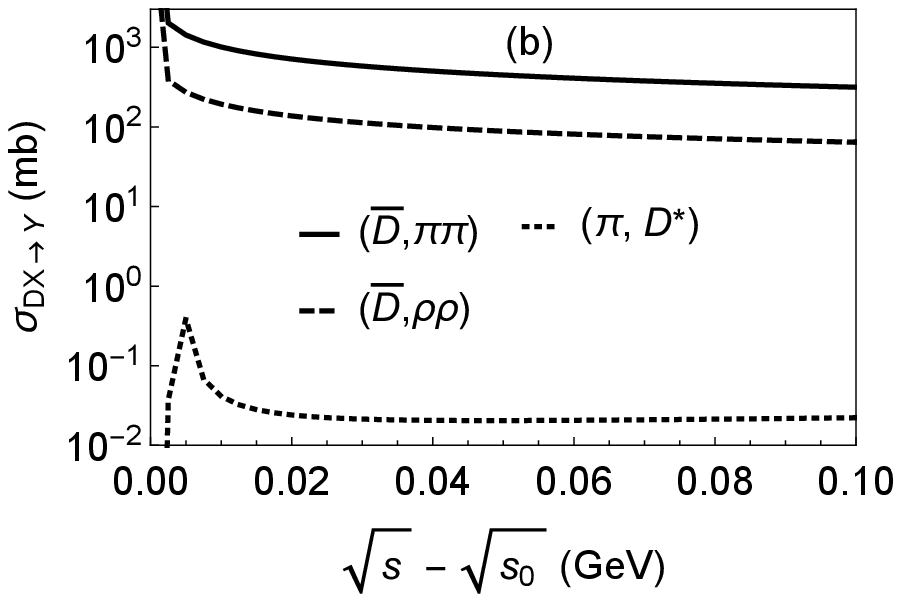}
\caption{ a) $D^*$ absorption cross sections for processes (1)-(5) shown in
Figs.~\ref{DIAG1} and~\ref{DIAG2} as a function of the CM energy $\sqrt{s}$.  
b) $D$ absorption cross sections for processes (6)-(7) shown in Fig.~\ref{DIAG2}
and for that described in Eq.~(\ref{CrSecFormX}) as a function of the CM energy
$\sqrt{s}$. The label $ (X,Y) $ identifies the respective channel
$D^{(*)} X \to Y $ considered. }
\label{CrSecDSTARD-ABS}
\end{figure}

In  Fig.~\ref{CrSecDSTARD-ABS}a the $D^*$ absorption cross sections for the 
processes summarized in Figs.~\ref{DIAG1} and~\ref{DIAG2} are plotted as a 
function of the CM energy $\sqrt{s}$.   It can be seen that the exothermic
processes with only light mesons in the final state have higher magnitudes at
smaller energies. Among these, the $D^* \bar{D}^* \to \pi \pi $ presents a
faster decreasing as the energy increases. On the other hand, up to moderate 
energies the process $D^* \bar{D}^* \to \rho \rho $ is larger than other
reactions, as a consequence of two effective couplings involving three vector 
mesons~\cite{suhoung}. In the case of reactions with a $D$ meson in the final
state, the only endothermic process, $D^* \pi \to \rho D $, has the cross       
section approximately  one order of magnitude higher than that of the reaction
$D^* \rho \to \pi D  $ for higher energies. In Fig.~\ref{CrSecDSTARD-ABS}b,    
the $D$ absorption cross sections are displayed. The processes with only light
mesons in the final state have greater cross sections typically by two or
three orders of magnitude.  

These results allow to quantitatively estimate and compare the different 
contributions for the $D^*$ and $D$ absorptions. Also,  they 
have smaller magnitudes when contrasted with the equivalent reactions
involving strange mesons reported in Refs.~\cite{suhoung,abreu,chiara21}. 

\begin{figure}[th]
\centering
\includegraphics[width=1.0\linewidth]{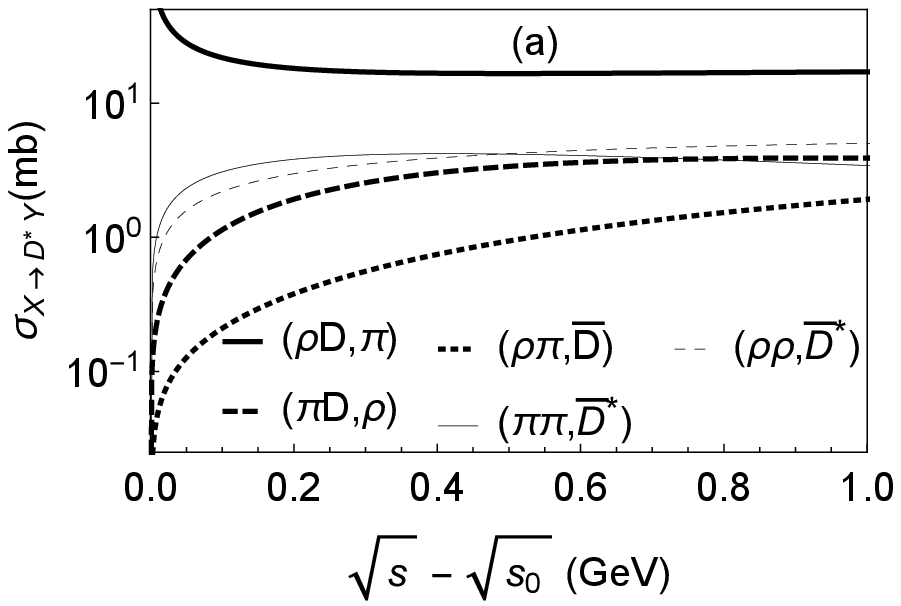}
\\
\vskip0.8cm
\includegraphics[width=1.0\linewidth]{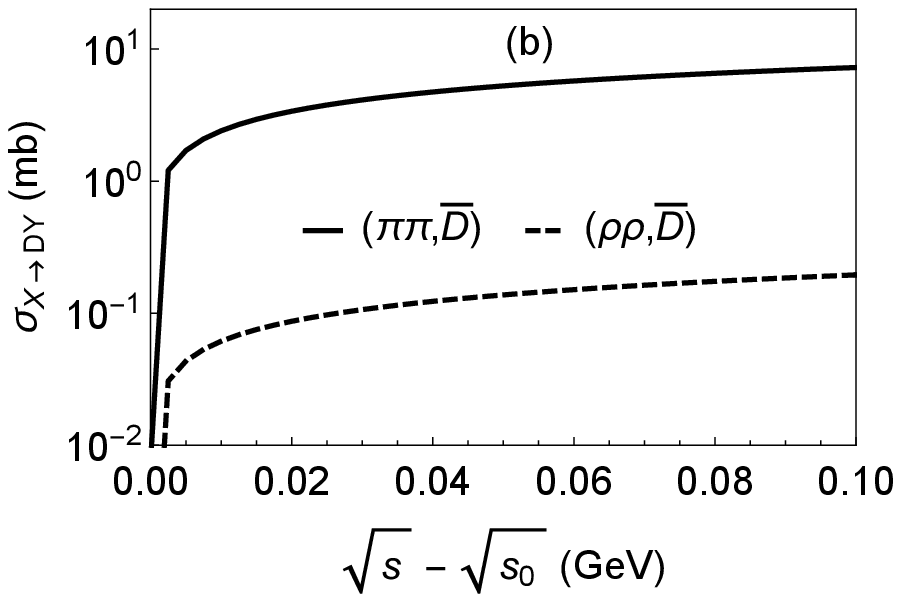}
\caption{ a) $D^*$ production cross sections for inverse processes 
    (1)-(5) shown in Figs.~\ref{DIAG1} and~\ref{DIAG2} as a function of the CM
    energy $\sqrt{s}$. b) $D$ production cross sections for inverse processes   
    (6)-(7) shown in Fig.~\ref{DIAG2} as a function of the CM energy $\sqrt{s}$. 
    The label $ (X,Y) $ identifies the respective channel $ X \to D^{(*)} Y $
    considered.}
\label{CrSecDSTARD-PROD}
\end{figure}

For the sake of completeness, we also present the cross sections related to  
the $D^*$ and $D$ production, obtained through the detailed balance relation 
involving the aforementioned processes. They are shown
in Fig.~\ref{CrSecDSTARD-PROD} as functions of the CM energy.           
Except for the case of $\sigma _{\rho D \rightarrow  D^* \pi } $, they are all
endothermic. The effects encoded in the detailed balance relations produce 
cross sections with different magnitudes when compared with the results shown   
in Fig.~\ref{CrSecDSTARD-ABS}.

To summarize: with a few exceptions, we find cross sections which are mostly
in the range $0.01 \, - \, 10$ mb, in rough agreement with most of the other
existing calculations \cite{crosses}. The exceptions are the charm
annihilation processes $D \bar{D} \to \pi \pi$, $D \bar{D} \to \rho \rho$
and $D^* \bar{D^*} \to \rho \rho$, which are much larger.
However, these processes do not contribute much
to the rate equation (see below) since their cross sections appear multiplied
by the square of the charm density ($n_D$ or $n_{\bar{D}}$), which is a small
number.

Having these cross sections, the next step is to compute the thermal
cross sections.

\subsection{Thermal cross sections}

In a relativistic heavy-ion collision  the reactions discussed above should  
occur typically in a later stage, namely in an equilibrated hadron medium at
temperatures between 100 and 160 MeV. Then, the relevant dynamical quantity  
is the cross section calculated in the kinematical regime where the colliding
particles have momenta of the order of the temperature. We define the
thermally averaged cross section, or simply thermal cross section, for a given
process $a b \rightarrow c d$ as:
\begin{eqnarray}
\langle \sigma_{a b \rightarrow c d } v_{a b}\rangle &  = & 
\frac{ \int  d^{3} \mathbf{p}_a  d^{3} \mathbf{p}_b
  \, f_a(\mathbf{p}_a) \,  f_b(\mathbf{p}_b) \,  \sigma_{a b \rightarrow c d } 
\,\,v_{a b} }{ \int d^{3} \mathbf{p}_a  
d^{3} \mathbf{p}_b \,  f_a(\mathbf{p}_a) \,  f_b(\mathbf{p}_b) }
\nonumber \\
& = & \frac{1}{4 \, \beta_a^2 K_{2}(\beta_a) \, \beta_b^2 \, K_{2}(\beta_b) } 
\nonumber \\
& & \times \int _{z_0} ^{\infty } dz \,  K_{1}(z) \,\,\sigma (s=z^2 T^2) 
\nonumber \\
& & \times \left[ z^2 - (\beta_a + \beta_b)^2 \right]
\left[ z^2 - (\beta_a - \beta_b)^2 \right],
\nonumber \\
  \label{tax}
\end{eqnarray}
where $v_{ab}$ is the relative velocity between the two initial interacting  
particles $a$ and $b$, the function $f_i(\mathbf{p}_i)$ is the temperature
$T$-dependent Bose-Einstein distribution of particles of species
$i$, $\beta _i = m_i / T$, $z_0 = max(\beta_a + \beta_b,\beta_c 
+ \beta_d)$, and $K_1$ and $K_2$ the modified Bessel functions of second kind.

\begin{figure}[th]  
\centering  
\includegraphics[width=1.0\linewidth]{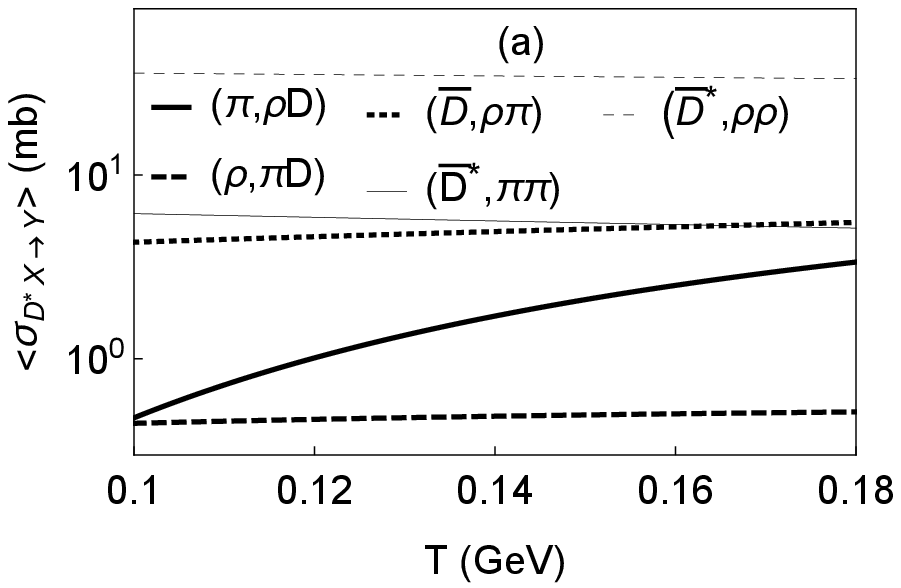}
\\
\vskip0.8cm
\includegraphics[width=1.0\linewidth]{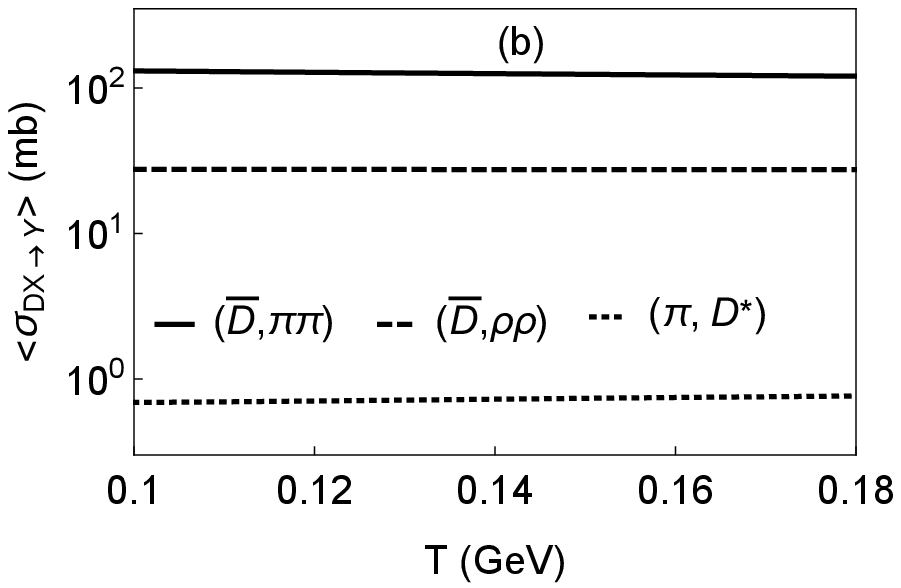}
\caption{a) Thermal cross sections as a function  of the temperature for the 
$D^*$  absorption via processes (1)-(5) shown in Figs.~\ref{DIAG1}
and~\ref{DIAG2}.
b) Same as a) but for the  $D$ absorption via processes (6)-(7) shown 
in Fig.~\ref{DIAG2} and for that described in Eq.~(\ref{CrSecFormX}).
The label $ (X,Y) $ identifies the respective channel $D^{(*)} X \to Y $
considered.}
\label{thermalcrsec1}  
\end{figure}  

\begin{figure}[th]  
\centering  
\includegraphics[width=1.0\linewidth]{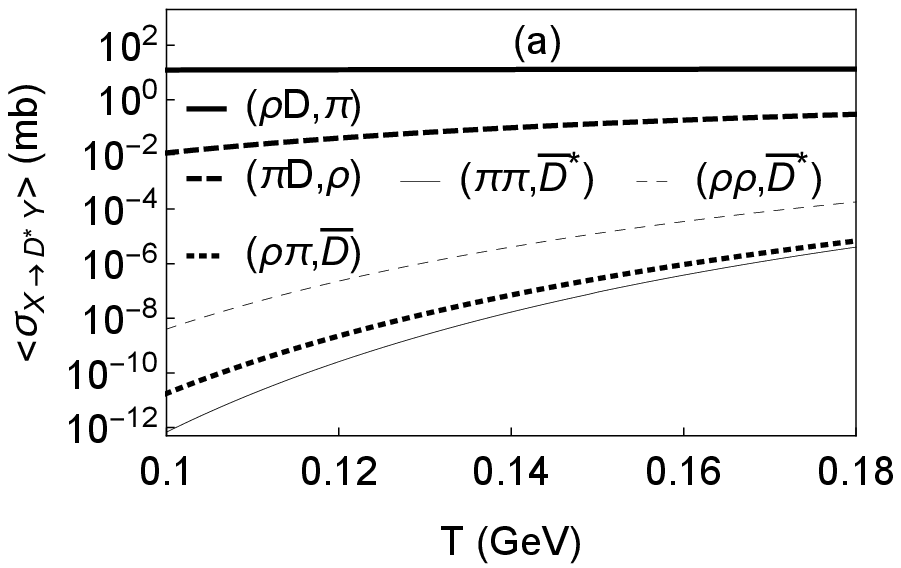} 
\\
\vskip0.8cm
\includegraphics[width=1.0\linewidth]{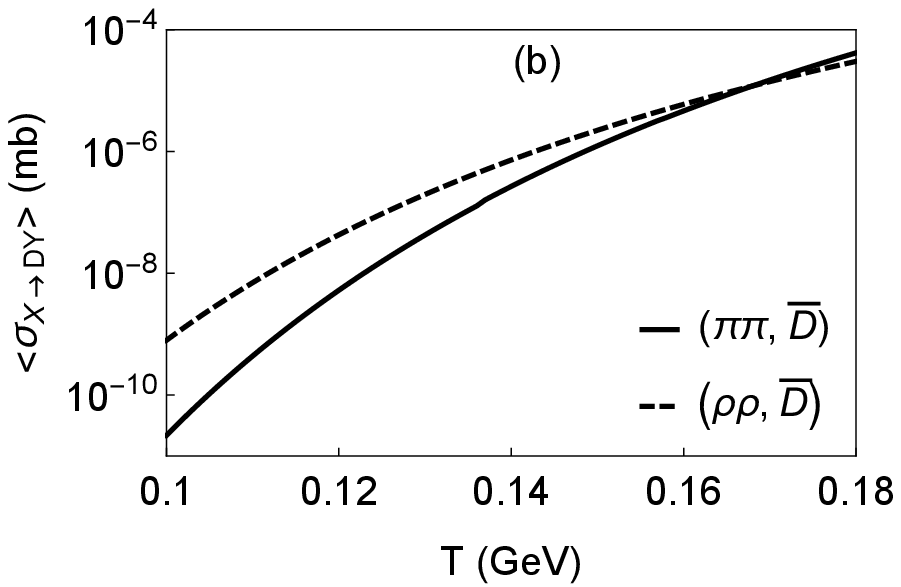}  
\caption{a) Thermal cross sections as a function of the temperature for the 
$D^*$  production via inverse processes to those shown in Figs.~\ref{DIAG1}
and~\ref{DIAG2}. 
b) Same as a) but for the $D$ production via inverse processes to those shown 
in Fig.~\ref{DIAG2}. The label $ (X,Y) $ identifies the respective channel
$X \to D^{(*)} Y $ considered.}
\label{thermalcrsec2}  
\end{figure}

In Figs.~\ref{thermalcrsec1} and~\ref{thermalcrsec2}  we plot the thermal      
cross sections as a function  of the temperature for the $D^*$ and $D$ meson 
absorption via processes (1)-(7) shown in Figs.~\ref{DIAG1} and~\ref{DIAG2}.  
For the sake of comparison, the thermal cross sections for the respective     
inverse reactions are also plotted. Within the considered range of temperatures, 
it can be seen that the reaction $ \rho D \to D^* \pi $ is the only  $D^*$   
production reaction with larger cross section than the corresponding inverse
reaction. In general, the thermal $D^*$ dissociation cross sections
are bigger than those for the production reactions. Also, we notice that
the $D^*$ meson absorption is easier by charmed mesons than by light mesons.
More specifically, the process $D^* \bar{D}^* \to \rho \rho $ has the largest
thermal cross section, being higher than the $D^*$ meson absorption by
$\pi, \rho$ mesons by one or two orders of magnitude. Interestingly, the
$D^{(*)}$ production reactions from light mesons are several orders of 
magnitude smaller than the corresponding dissociation processes, reflecting
the higher energy  threshold  of these processes.

To conclude this section, we highlight the differences among the magnitudes
of thermal cross sections for  $D^{(*)}$  dissociation and production  
reactions, which in principle might play an important role in the time
evolution of the  $D^*$ multiplicity.

\section{Evolution equations} 
\label{EvolEq} 

Now we move to the analysis of the time evolution of the $D^*$ and $D$        
multiplicities during the hadronic stage of heavy ion collisions. To this end, 
the thermal  cross sections estimated in the previous section will be used as 
input in the momentum-integrated evolution equations for the $D^*$ and $D$     
abundances to estimate the gain and loss terms due to $D^{(*)}$ production and  
absorption. As in the previous works~\cite{suhoung,chiara21}, these equations 
with all considered creation and annihilation reactions are given by: 
\begin{widetext}
  \begin{eqnarray}
  	\frac{dN_{D^*}}{d \tau} & = & \braket{\sigma_{D \rho \rightarrow D^* \pi} v_{D \rho}}n_{\rho}(\tau)N_D(\tau)-\braket{\sigma_{D^* \pi \rightarrow D \rho} v_{D^* \pi}} n_\pi (\tau) N_{D^*} (\tau) + \braket{\sigma_{D \pi \rightarrow D^* \rho} v_{D \pi}}n_{\pi}(\tau)N_D(\tau) \notag
\\ &  &- \braket{\sigma_{D^* \rho \rightarrow D \pi} v_{D^* \rho}}n_{\rho}(\tau)N_{D^*}(\tau) + \braket{\sigma_{\pi \rho \rightarrow D^* \bar{D}} v_{\pi \rho}}n_{\pi}(\tau)N_\rho(\tau) - \braket{\sigma_{D^* \bar{D} \rightarrow \rho \pi} v_{D^* \bar{D}}}n_{\bar{D}}(\tau)N_{D^*}(\tau) \notag 
\\ & &+ \braket{\sigma_{\pi \pi \rightarrow D^* \bar{D}^*} v_{\pi \pi}}n_{\pi}(\tau)N_\pi(\tau)-\braket{\sigma_{D^* \bar{D}^* \rightarrow \pi \pi} v_{D^* \bar{D}^*}}n_{\bar{D}^*}(\tau)N_{D^*}(\tau) + \braket{\sigma_{\rho \rho \rightarrow D^* \bar{D}^*} v_{\rho \rho}}n_{\rho}(\tau)N_\rho(\tau) \notag
	\\ & &- \braket{\sigma_{D^* \bar{D}^* \rightarrow \rho \rho} v_{D^* \bar{D}^*}}n_{\bar{D}^*}(\tau)N_{D^*}(\tau) + \braket{\sigma_{D \pi \rightarrow D^*} v_{D \pi}}n_{\pi}(\tau)N_D(\tau) - \braket{\Gamma_{D^*}} N_{D^*}(\tau), \nonumber	
  \end{eqnarray}
  \begin{eqnarray}
	\frac{dN_{D}}{d \tau} & = & \braket{\sigma_{\pi \pi \rightarrow D \bar{D}} v_{\pi \pi}}n_{\pi}(\tau)N_\pi(\tau)-\braket{\sigma_{D \bar{D} \rightarrow \pi \pi} v_{D \bar{D}}}n_{\bar{D}}(\tau)N_{D}(\tau)+\braket{\sigma_{\rho \rho \rightarrow D \bar{D}} v_{\rho \rho}}n_{\rho}(\tau)N_\rho(\tau) \notag \\
	& &-\braket{\sigma_{D \bar{D} \rightarrow \rho \rho} v_{D \bar{D}}}n_{\bar{D}}(\tau)N_{D}(\tau)+\braket{\sigma_{D^* \pi \rightarrow D \rho} v_{D^* \pi}} n_\pi (\tau) N_{D^*} (\tau) - \braket{\sigma_{D \rho \rightarrow D^* \pi} v_{D \rho}}n_{\rho}(\tau)N_D(\tau) \notag \\
	& &+ \braket{\sigma_{D^* \rho \rightarrow D \pi} v_{D^* \rho}}n_{\rho}(\tau)N_{D^*}(\tau)-\braket{\sigma_{D \pi \rightarrow D^* \rho} v_{D \pi}}n_{\pi}(\tau)N_D(\tau) +\braket{\sigma_{\pi \rho \rightarrow D^* \bar{D}} v_{\pi \rho}}n_{\pi}(\tau)N_\rho(\tau) \notag \\
	& &-\braket{\sigma_{D^* \bar{D} \rightarrow \rho \pi} v_{D^* \bar{D}}}n_{\bar{D}}(\tau)N_{D^*}(\tau)+\braket{\Gamma_{D^*}} N_{D^*}(\tau)-\braket{\sigma_{D \pi \rightarrow D^*} v_{D \pi}}n_{\pi}(\tau)N_D(\tau) .
	\label{rateeqs}
\end{eqnarray} 
\end{widetext}
where $n_{i} (\tau)$ are $N_{i}(\tau)$ denote the density and the abundances
of the involved mesons  at proper time $\tau$. 

The rate equations above and their gain and loss contributions deserve some    
comments. The decay of $D^*$ and its regeneration from the daughter particles 
$ D $ and $\pi$ are included in the last line of the two evolution equations.
However, due to the small value of the $D^*$ decay width $\Gamma_{D^*}$, the 
lifetime of the $D^*$ mesons is much greater than that of the hadron gas
presumed in this  work (of the order of 10 fm/c). Therefore, the decay of 
$D^*$ can be neglected in the rate equation. Moreover, as in
Refs.~\cite{suhoung,abreu,chiara21}, some terms yield very small 
contributions because of the smallness of their thermal cross sections
and/or small values of $D^{(*)}$ densities with respect to the ones of
light mesons, which are the most abundant particles in a hadron gas.
Then, several terms associated to the processes with two charmed mesons
in initial or final states can be safely neglected.

To obtain the solutions of Eq.~(\ref{rateeqs}) we assume that the pions, 
rho and charm  mesons in the reactions contributing to the $D^*$ and $D$    
multiplicities are in thermal equilibrium. Then, the density $n_{i} (\tau)$ 
is written as
~\cite{Chen:2007zp,Abreu:2018mnc,Abreu:2020ony,Koch,ChoLee1,Cho:2017dcy}
\begin{eqnarray}
  n_{i} (\tau) &  \approx & \frac{1}{2 \pi^2} \, \gamma_{i} \, g_{i} \,
  m_{i}^2 \,  T(\tau) \, K_{2}\left(\frac{m_{i} }{T(\tau)}\right), 
\label{densities}
\end{eqnarray}
where $\gamma _i$ and $g_i$ are the fugacity factor and the  degeneracy
factor of the particle of type $i$, respectively. We obtain the multiplicity
$N_i (\tau)$ multiplying the density $n_i(\tau)$ by the volume $V(\tau)$.  
To model the dynamics of relativistic heavy ion collisions in the hadronic 
phase, the temperature $T(\tau)$ and volume $V(\tau)$ are parametrized 
according to the boost invariant Bjorken picture with an accelerated 
transverse expansion~\cite{Abreu:2018mnc,Abreu:2020ony,ChoLee1,Cho:2017dcy}.
The $\tau$ dependence of $V(\tau)$ and $T$ are thus given by
  \begin{eqnarray}
    T(\tau) & = & T_C - \left( T_H - T_F \right) \left( \frac{\tau - \tau _H }
{\tau _F -  \tau _H}\right)^{\frac{4}{5}} , \nonumber \\
V(\tau) & = & \pi \left[ R_C + v_C 
\left(\tau - \tau_C\right) + \frac{a_C}{2} \left(\tau - \tau_C\right)^2 
\right]^2 \tau \, c , \nonumber \\
\label{TempVol}
  \end{eqnarray}
where $R_C $ and $\tau_C$  denote the final transverse and longitudinal   
sizes of the QGP; $v_C $ and  $a_C $ are its transverse flow velocity and
transverse acceleration at $\tau_C $; $T_C$ is the critical temperature  
of the quark-hadron transition; $T_H $ is the temperature of the hadronic 
matter at the end of the mixed phase, occurring at the time $\tau_H $;    
and the kinetic freeze-out temperature  $T_F $ leads to a freeze-out time 
$\tau _F $. We notice that this simple picture of the hydrodynamic 
evolution of fluid should be seen as a tool to mimic the           
essential features of hydrodynamic expansion and cooling of the hadron gas.  
The collisions  chosen for this study are the central
$Pb-Pb$ collisions at $\sqrt{s_{NN}} = 5$ TeV at the LHC. In this way, we
use the parameter choice of Ref.~\cite{Cho:2017dcy}, which, for convenience,
is summarized in Table~\ref{parameters}.

\begin{center}
\begin{table}[h!]
  \caption{Parameters used in Eq.~(\ref{TempVol}) for central
    $Pb-Pb$ collisions at $\sqrt{s_{NN}} = 5$ TeV ~\cite{Cho:2017dcy}.}
\vskip1.5mm
\label{parameters}
\begin{tabular}{ c c c }
\hline
\hline
 $v_C$ (c) & $a_C$ (c$^2$/fm) & $R_C$ (fm)   \\   
0.5 & 0.09 & 11  
\\  
\hline
 $\tau_C$ (fm/c) & $\tau_H$ (fm/c)  &  $\tau_F$ (fm/c)  \\   
7.1  & 10.2 & 21.5
\\  
\hline
  $T_C (\MeV)$  & $T_H (\MeV)$ & $T_F (\MeV)$ \\   
 156 & 156 & 115   \\  
\hline
 $N_c$  & $N_{\pi}(\tau_F)$ & $N_{\rho}(\tau_F)$ \\   
 14 & 2410 & 179 \\  
\hline
 $N_{D}(\tau_H)$  & $N_{D^*}(\tau_H)$ & \\   
 4.7 & 6.3 &  \\  
\hline
\hline
\end{tabular}
\end{table}
\end{center}

We assume that the total number of charm quarks $(N_c)$ in charmed hadrons
remains approximately conserved during the fireball evolution,  i .e.      
$n_c(\tau) \times V(\tau) = N_c$. Then we use a time-dependent charm quark
fugacity factor $\gamma _c $ in Eq.~(\ref{densities}) to calculate the     
charmed mesons in order to keep $N_c$ constant. Besides, the total number 
of  pions and $\rho$ mesons at freeze-out was taken from 
Refs.~\cite{ChoLee1,Abreu:2020ony}. These numbers are displayed in
Table~\ref{parameters}. 

\begin{figure}[!ht]
\includegraphics[width=1.0\linewidth]{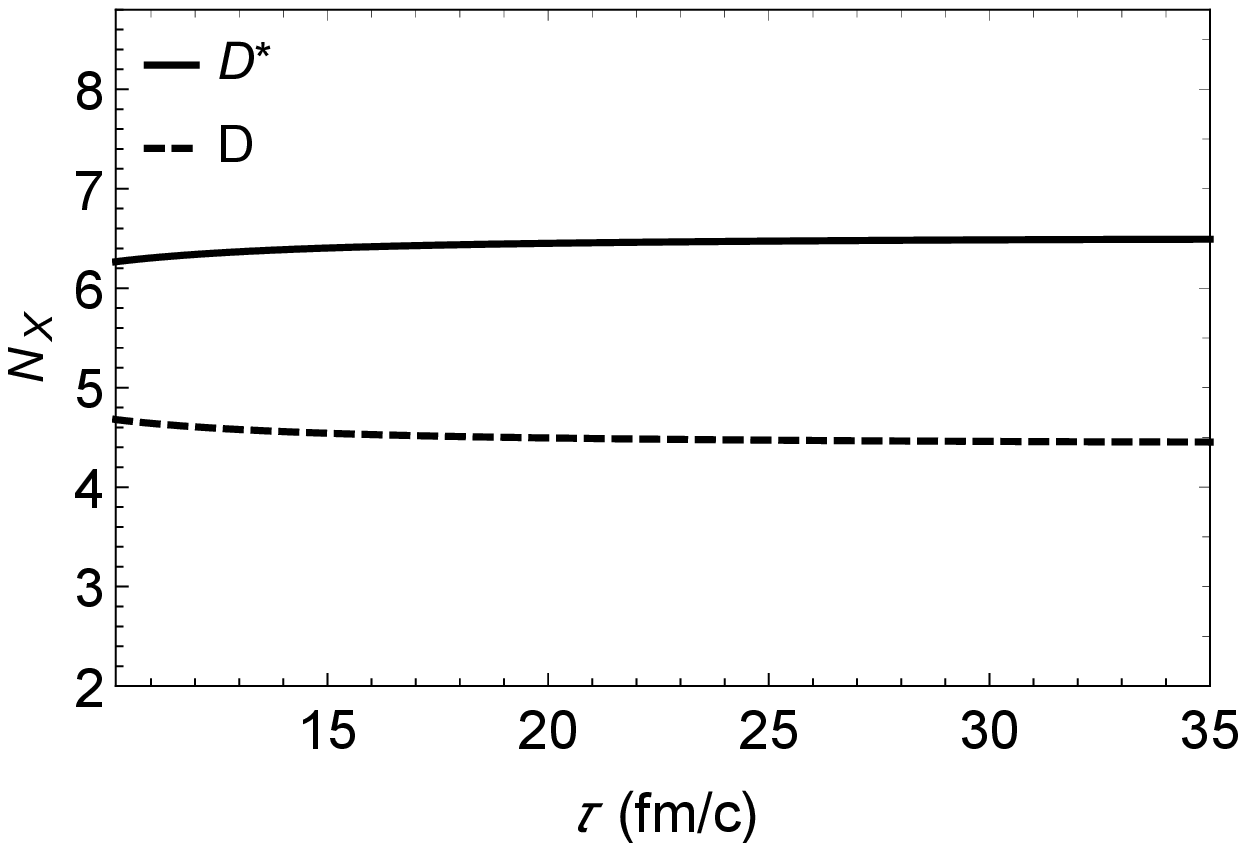}
\vskip0.8cm
\includegraphics[width=1.0\linewidth]{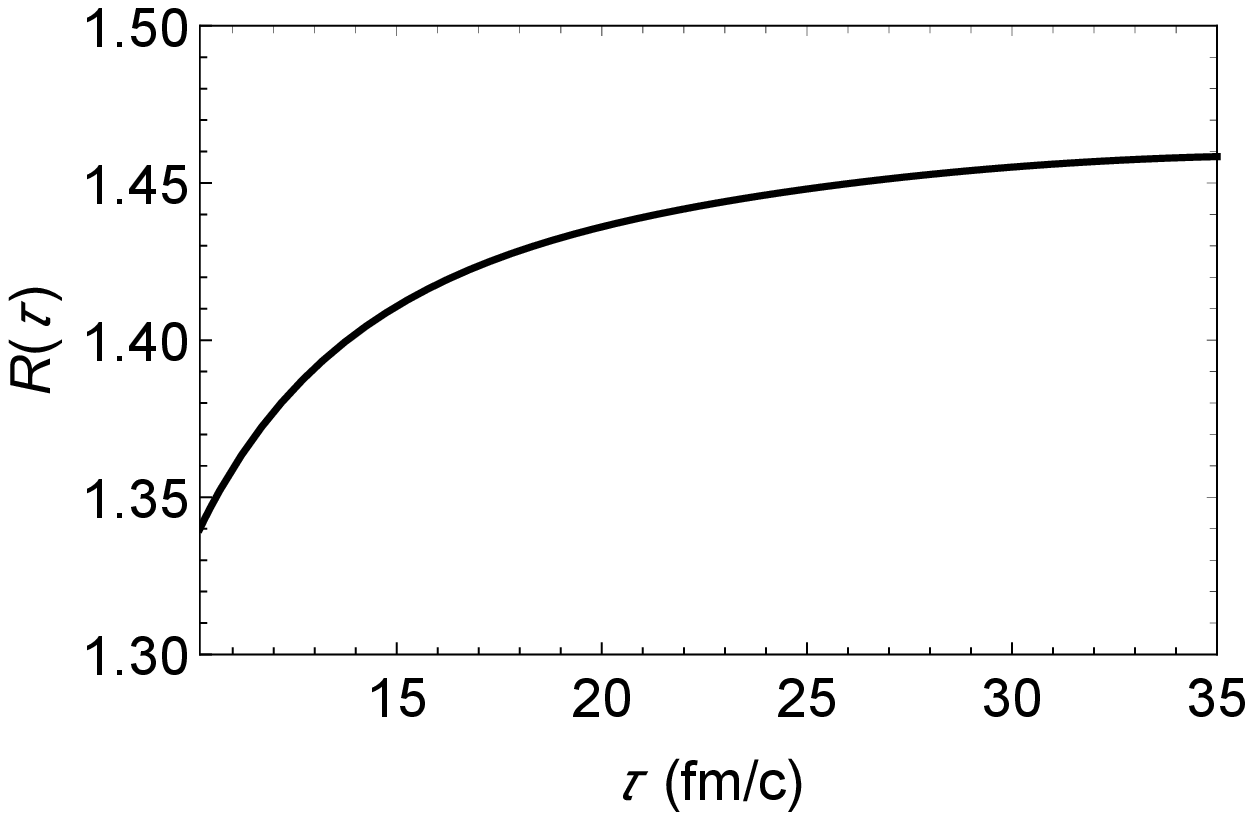}
\caption{Upper panel:
  $D$ and $D^*$ abundances as a function of the proper time $\tau$ in
  central $Pb-Pb$  collisions at $\sqrt{s_{NN}} = 5$ TeV, calculated with
  the parameters from Table~\ref{parameters}. Lower panel: 
  evolution of the ratio $ R= {D^*}/{D} $.}
\label{TimeEvolDDSTAR}
\end{figure}

The time evolution of the $D$ and $D^*$ 
abundances is plotted in Fig.~\ref{TimeEvolDDSTAR} as a function of  
the proper time.  These results suggest that the interactions of the $D$'s
and $D^*$ with the hadronic medium  through the reactions discussed
previously in HICs produce only small changes in the multiplicities:
while $N_{D^*}$ grows $5 \% $, $N_{D}$ decreases by $7 \% $. In the figure
we also show the ratio $ R= {D^*}/{D} $.  This is the main result of our
work. We conclude that the information contained in open charm multiplicities
will be propagated from the hadronization time to the kinetic freeze-out time
without much distortion.

%

Finally, in order to make predictions wich can be confronted with 
experimental data we compute the ratio $R$ as a function  of the         
multiplicity density of charged particles measured at midrapidity, i.e.
$\mathcal{N} = \left[ d N_{ch} / d \eta (\eta < 0.5)\right]^{1/3}$.
The quantity $\mathcal{N}$ is usually considered a measure of the size
of the  system~\cite{Aamodt:2011mr,Lisa:2005dd}, which in turn is directly 
related to the kinetic freeze-out temperature~\cite{alice13}. An empirical  
connection between these variables was established in Ref.~\cite{chiara21}. 
Assuming that the hadron gas suffers a Bjorken-like cooling, the            
freeze-out time $\tau_f$ is related to the freeze-out temperature $T_f$ through
the expression: 
\begin{equation} 
\tau_f = \tau_h  \left( \frac{T_H}{T_F} \right)^3.
\label{bjorf}
\end{equation}
Next, we use the empirical relation between $T_F$ and
$\mathcal{N} $ extracted from ~\cite{alice13} and parametrized
as ~\cite{chiara21}: 
\begin{equation}
T_F  = {T_{F0}} \, e^{- b \, \mathcal{N}},
\label{chiafit}
\end{equation}
where $T_{F0} = 132.5$ MeV and $ b = 0.02$.
Then, inserting (\ref{chiafit}) into (\ref{bjorf}) we obtain
\beq
\tau_F \propto e^{3 b \mathcal{N}}.
\label{relf}
\eeq
Thus, $\mathcal{N}$ provides an estimate of the duration of the
hadronic phase. Larger systems have bigger $\mathcal{N}$ and live longer. 
We use Eq.~(\ref{relf}) in the solutions of Eq.~(\ref{rateeqs}) to
determine $R$ as a function  $\mathcal{N}$. 

\begin{figure}[!ht]
\begin{center}
      \includegraphics[width=1.0\linewidth]{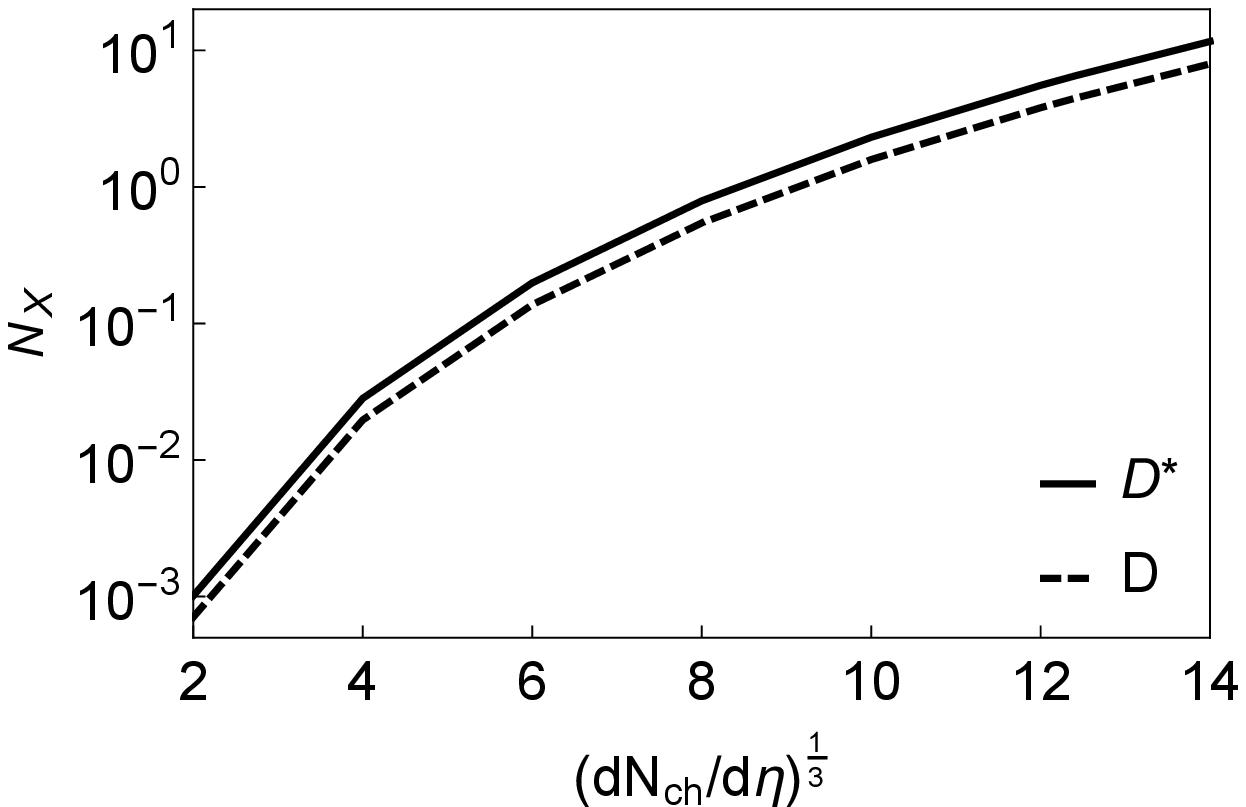} \\
      \vskip0.8cm
      \includegraphics[width=1.0\linewidth]{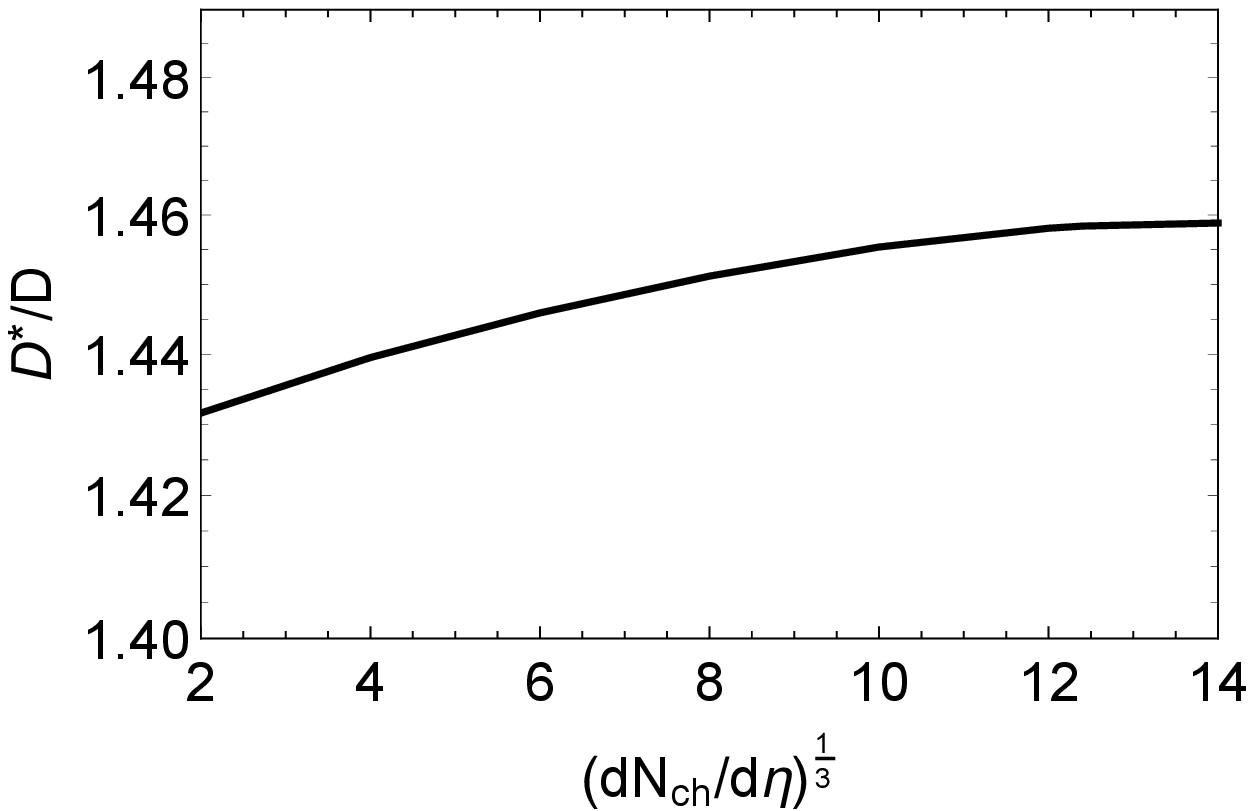}
\end{center}
\caption{Upper panel: $D$ and $D^*$ abundances as a function of $\mathcal{N}$.
  Lower panel: $D^* / D$  with initial condition $D^* / D = 1.34 $  as a
  function of  $\mathcal{N}$. }
\label{radata}
\end{figure}

In Fig. \ref{radata} we show the $D^*$ and $D$ abundances and the ratio
$R = D^* / D$ as a function of $\mathcal{N}$.
As it can be seen in the upper panel of the figure, the abundances depend 
strongly on the system size. However,  the ratio $D^* / D$ does not change 
significantly. This fact contrasts with the case of the strange $K^*$ and
$K$ mesons discussed in Ref.~\cite{chiara21}, where  the longer the hadronic
system lasts, the smaller is the ratio $K^*/K$. This difference can be
attributed to the difference in the decay widths $D^* \to D \, \pi $ and
$K^* \to K \, \pi $, with the latter being much larger.  

To the best of our knowledge, there are so far no data available  to 
perform a direct comparison with the predictions presented in 
Fig.~\ref{radata}.

\section{Concluding Remarks} 
\label{Conclusions} 

In this work we have performed a systematic analysis of the $D^{(*)}$ meson
dissociation in the hadron gas phase of heavy ion collisions.
Using an  effective Lagrangian formalism, we have evaluated the thermal
production and absorption  cross sections of the $D^*$ and $D$ mesons in a
hadron gas. The couplings and form factors in all the vertices were
previously computed  with QCD sum rules. Therefore, in our case, much 
of the uncertainties and arbitrariness, inherent to this kind of
calculation, were avoided.

The obtained cross sections were used as input in the rate 
equations to estimate the time evolution of the  $D^{(*)}$ multiplicities
and the ratio $D^* /D$ during the hadron gas phase of heavy ion collisions.  
The multiplicities of $D$ and $D^*$ remain almost constant during the evolution
of the hadron gas. The ratio $D^* / D $ suffers only a slight increase. 
This result indicates that charm is already in chemical equilibrium at the
beginning of the hadron gas phase. Therefore our calculation can be regarded as
a microscopic justification of the use of the statistical hadronization model
to compute the charm meson abundances.

The lifetime of a hadronic fireball is, of course, not directly measurable and
we would like to make contact with expreimental data. To this end, we have
assumed a Bjorken type cooling for the hadron gas and used an empirical     
relation between the freeze-out temperature and the multiplicity density of
charged particles at midrapidity. With this procedure  we estimated the 
ratio $D^* /D$ as a function of ($ dN /d \eta (\eta =0)$). As expected from 
the first part of the calculation, we found that $D^* /D$ slightly increases
if the hadronic medium lives longer, in sharp contrast to the $K^* /K$ ratio,
computed in \cite{chiara21} with the same methods.

We believe that these results can be compared with future experimental data,
and may contribute to a better understanding  of the hadron medium created
in heavy ion collisions.

\begin{acknowledgements}

The authors would like to thank the Brazilian funding agencies for their 
financial support: CNPq (LMA: contracts 309950/2020-1 and 400546/2016-7), 
FAPESB (LMA: contract INT0007/2016) and the INCT-FNA.

\end{acknowledgements} 

\appendix

\section{Amplitudes}
\label{Ampl}

Here we give the explicit expressions for the contributions to the amplitudes in Eq.~(\ref{Amplitudes}), associated to the processes shown in Figs.~\ref{DIAG1} and.~\ref{DIAG2}. They are~\cite{suhoung,abreu,chiara21}
\begin{widetext}
  \begin{eqnarray}
	\mathcal{M}_{(1a)} & = & \tau _{rs} ^{(i)} \tau _{r's'} ^{(j)}  g_{\pi D D^* }  g_{\rho D^* D^*}  \epsilon^{\alpha}_{1} \epsilon^{*\beta}_{3} \frac{1}{t-m^{2}_{D^*} + im_{D^*} \Gamma_{D^*}} \left( -g^{\mu \nu} + \frac{(p_1 - p_3)^{\mu} (p_1 - p_3)^{\nu}}{m^{2}_{D^*}} \right) (p_2 + p_4)_{\mu}
	\nonumber \\
& & \times \left( (2 p_1 - p_3)_{\beta g_{\alpha \nu}} - (p_1 + p_3)_{\nu g_{\alpha \beta}} - (p_1 - 2 p_3)_{\alpha g_{\beta \nu}} \right),
	\nonumber \\
	\mathcal{M}_{(1b)} & = & -\tau _{rs} ^{(i)} \tau _{r's'} ^{(j)}  g_{\pi D D^* }  g_{\rho D D} \epsilon^{\mu}_{1} \epsilon^{* \nu}_{3} \frac{1}{s -  m^2_D} (p_1 + 2 p_2)_{\mu} (p_3 + 2p_4)_{\nu}, \nonumber \\
 \mathcal{M}^{ (1.c) } & = & g_{\rho D D^{*}} g_{\pi D^{*} D^{*}} \varepsilon_{1}^{\mu} \varepsilon_{3}^{\nu} \epsilon_{\mu \gamma \delta \alpha}    \epsilon_{\nu \sigma \rho \beta }  \frac{1}{s – m_{D^{*}}^{2} + i m_{D^{*}} \Gamma_{D^{*}}} 
	 \left( -g^{\alpha \beta} + \frac{ (p_1 + p_2)^{\alpha} (p_1 + p_2)^{\beta} }{ m_{D^{*}}^{2}} \right ) p_{1}^{\gamma} p_{2}^{\delta} p_{3}^{\sigma} p_{4}^{\rho}, 
\label{M1}  
\end{eqnarray} 
 
  \begin{eqnarray} 
	\mathcal{M}_{(2a)} & = & - \tau _{rs} ^{(i)} \tau _{r's'} ^{(j)} g_{\pi D D^*} g_{\rho D D} \epsilon^{\mu}_{1} \epsilon^{\nu}_{2} \frac{1}{t - m^2_D} (p_1 - 2p_3)_{\mu} (2p_4 - p_2)_{\nu},  \nonumber \\
\nonumber	\mathcal{M}_{(2b)} & = &  - \tau _{rs} ^{(i)} \tau _{r's'} ^{(j)} g_{\pi D D^* } g_{\rho D^* D^*} \epsilon^{\alpha}_{1} \epsilon^{\beta}_{2} \frac{1}{s - m^2_{D^*} + im_{D^*} \Gamma_{D^*}} \left(-g^{\mu \nu} + \frac{(p_1 + p_2)^{\mu} (p_1 + p_2)^{\nu}}{m^2_{D^*}} \right) (p_3 - p_4)_{\mu}, \\
& &  \times \left( (2p_1 + p_2)_{\beta g_{\alpha \nu}} - (p_1 - p_2)_{\nu g_{\alpha \beta}} - (p_1 + 2p_2)_{\alpha g_{\beta \nu}} \right)
\nonumber \\
 \mathcal{M}^{ (2.c) } & = & g_{\rho D D^{*}} g_{\pi D^{*} D^{*}} \varepsilon_{1}^{\mu} \varepsilon_{2}^{\nu} \epsilon_{\mu \gamma \delta \alpha}    \epsilon_{\nu \sigma \rho \beta }  \frac{1}{t – m_{D^{*}}^{2} + i m_{D^{*}} \Gamma_{D^{*}}} 
	\left( -g^{\alpha \beta} + \frac{ (p_1 - p_3)^{\alpha} (p_1 - p_3)^{\beta} }{ m_{D^{*}}^{2}} \right ) p_{1}^{\gamma} p_{3}^{\delta} p_{2}^{\sigma} p_{4}^{\rho},
\label{M2} 
\end{eqnarray} 

	\begin{eqnarray} 
 \nonumber	\mathcal{M}_{(3a)} & = & \tau _{rs} ^{(i)} \tau _{r's'} ^{(j)} g_{\pi D D^* } g_{\rho D^* D^*} \epsilon^{\alpha}_{1} \epsilon^{*\beta}_{3} \frac{1}{t- m^{2}_{D^*} + im{D^*} \Gamma_{D^*}} \left( -g^{\mu \nu} + \frac{(p_1 - p_3)^{\mu} (p_1 - p_3)^{\nu}}{m^{2}_{D^*}} \right) (p_2 + p_4)_{\mu} \\
& & \times \left( (2p_3 - p_1)_{\alpha g_{\beta \nu}} - (p_1 + p_3)_{\nu g_{\alpha \beta}} + (2p_1 - p_3)_{\beta g_{\alpha \nu}} \right),
  \nonumber \\
  	\mathcal{M}_{(3b)} & = & \tau _{rs} ^{(i)} \tau _{r's'} ^{(j)} g_{\pi D D^* } g_{\rho D D} \epsilon^{\mu}_{1} \epsilon^{* \nu}_{3} \frac{1}{u - m^2_D} (2p_4 - p_1)_{\mu} (2p_2 - p_3)_{\nu},
\nonumber \\
\mathcal{M}^{ (3.c) } & = & g_{\pi D^{*} D^{*}} g_{\rho D D^{*}} \varepsilon_{1}^{\mu} \varepsilon_{3}^{\nu} \epsilon_{\mu \gamma \delta \beta}  \epsilon_{\nu \sigma \rho \alpha}  \frac{1}{u – m_{D^{*}}^{2} + i m_{D^{*}} \Gamma_{D^{*}}}
	\left( –g^{\alpha \beta} + \frac{ (p_1 – p_4)^{\alpha} (p_1 – p_4)^{\beta} }{m-{D^{*}}^{2}} \right)  p_{1}^{\gamma} p_{4}^{\delta} p_{2}^{\rho} p_{3}^{\sigma},
\label{M3} 
\end{eqnarray} 
 
	\begin{eqnarray}
 	\mathcal{M}_{(4a)} & = & \tau _{rs} ^{(i)} \tau _{r's'} ^{(j)} g^{2}_{\pi D D^*} \epsilon^{\mu}_{1} \epsilon^{\nu}_{2} \frac{1}{t - m^2_D} (p_1 - p_3)_{\mu} (p_2 - 2p_4)_{\nu},
 \nonumber \\
 	\mathcal{M}_{(4b)} & = &  \tau _{rs} ^{(i)} \tau _{r's'} ^{(j)} g^{2}_{\pi D D^* } \epsilon^{\mu}_{1} \epsilon^{\nu}_{2} \frac{1}{u - m^2_D} (2p_1 - 2p_4)_{\mu} (p_2 - 2p_3)_{\nu}
 \nonumber \\
\nonumber \mathcal{M}^{ (4.c) } & =  & g_{\pi D^* D^* }^{2} \varepsilon_{1}^{\mu} \varepsilon_{2}^{\nu} \epsilon_{\mu \gamma \delta \alpha} \epsilon_{\nu \sigma \rho \beta} \frac{1}{t – m_{D^{*}}^{2} + i m_{D^{*}} \Gamma_{D^{*}}} 
	\left( -g^{\alpha \beta} + \frac{ (p_1 – p_3)^{\alpha} (p_1 – p_3)^{\beta} }{m_{D^{*}}^{2}} \right)   p_{1}^{\gamma} p_{3}^{\delta} p_{2}^{\sigma} p_{4}^{\rho},
	\nonumber \\
\mathcal{M}^{ (4.d) } & = & -  g_{\pi D^* D^* }^{2} \varepsilon_{1}^{\mu} \varepsilon_{2}^{\nu} \epsilon_{\mu \gamma \delta \alpha} \epsilon_{\nu \sigma \rho \beta} \frac{1}{u – m_{D^{*}}^{2} + i m_{D^{*}} \Gamma_{D^{*}}} 
	\left( -g^{\alpha \beta} + \frac{ (p_1 – p_4)^{\alpha} (p_1 – p_4)^{\beta} }{m_{D^{*}}^{2}} \right)   p_{1}^{\delta} p_{4}^{\gamma} p_{2}^{\sigma} p_{3}^{\rho},
\label{M4} 
\end{eqnarray} 
   
  \begin{eqnarray}
	\mathcal{M}_{(5a)} & = & \tau _{rs} ^{(i)} \tau _{r's'} ^{(j)} g^{2}_{\rho D^* D^*} \epsilon^{\alpha}_{1} \epsilon^{*\beta}_{3} \epsilon^{\gamma}_{2} \epsilon^{* \delta}_{4} \frac{1}{t - m^{2}_{D^*} + im_{D^*}\Gamma_{D^*}} \left( -g^{\mu \nu} + \frac{(p_1 - p_3)^{\mu} (p_1 - p_3)^{\nu}}{m^{2}_{D^*}} \right) \nonumber \\
& & \times \left((2p_3 - p_1)_{\alpha g_{\beta \mu}} - (p_1 + p_3)_{\mu g_{\alpha \beta}} + (2p_1 - p_3)_{\beta g _{\alpha \nu}} \right) \left( (p_2 + p_4)_{\gamma g_{\delta \nu}} + (p_2 - 2p_4)_{\nu g_{\gamma \nu}} \right),
 \nonumber \\
 \nonumber	\mathcal{M}_{(5b)} & = & \tau _{rs} ^{(i)} \tau _{r's'} ^{(j)} g^{2}_{\rho D^* D^*} \epsilon^{\alpha}_{1} \epsilon^{*\beta}_{4} \epsilon^{\gamma}_{2} \epsilon^{*\delta}_{3} \frac{1}{u - m^{2}_{D^*} + im_{D^*} \Delta_{D^*}} \left( -g^{\mu \nu} + \frac{(p_1 - p_4)^{\mu} (p_1 - P_4)^{\nu}}{m^{2}_{D^*}} \right)\\
& & \times \left( (2p_4 - p_1)_{\alpha g_{\beta \mu}} - (p_1 + p_4)_{\mu g_{\alpha \beta}} + (2p_1 - p_4)_{\beta g_{\alpha \nu}} \right) \left( (p_2 + p_3)_{\gamma g_{\gamma \nu}} + (p_2 - 2p_3)_{\nu g_{\gamma \nu}} + (p_3 - 2p_2)_{\delta g_{\gamma \nu}} \right), 
\nonumber \\
\mathcal{M}^{ (5.c) } & = & - g_{\rho D D^{*}}^{2} \varepsilon_{1}^{\alpha}  \varepsilon_{3}^{\beta} \varepsilon_{2}^{\gamma *} \varepsilon_{4}^{\delta *} \epsilon_{\alpha \beta \mu \nu}  \epsilon_{\gamma \delta \sigma \omega} \frac{1}{t – m_{D}^{2}} 
	p_{1}^{\nu} p_{3}^{\mu} p_{2}^{\sigma} p_{4}^{\omega},
	\nonumber \\
\nonumber \mathcal{M}^{ (5.d) } & = & g_{\rho D D^{*}}^{2} \varepsilon_{1}^{\alpha} \varepsilon_{3}^{\beta} \varepsilon_{2}^{\gamma} \varepsilon_{4}^{\delta} \epsilon_{\alpha \delta \mu \nu}  \epsilon_{\gamma \beta \sigma \omega} \frac{1}{u – m_{D}^{2}} 
	p_{1}^{\mu} p_{4}^{\nu} p_{2}^{\sigma} p_{3}^{\omega},
\label{M5} 
\end{eqnarray}
 
	\begin{eqnarray}
 	\mathcal{M}_{(6a)} & = & \tau _{rs} ^{(i)} \tau _{r's'} ^{(j)} g^{2}_{\pi D D^*} \frac{1}{t - m^{2}_{D^*} + im_{D^*}\Gamma_{D^*}} \left( -g^{\mu \nu} + \frac{(p_1 - p_3)^{\mu} (p_1 - p_3)^{\nu}}{m^{2}_{D^*}}  \right) (p_1 + p_3)_{\mu } (p_2 + p_4)_{\nu }, \nonumber \\
 	\mathcal{M}_{(6b)} & = & \tau _{rs} ^{(i)} \tau _{r's'} ^{(j)} g^{2}_{\pi D D^*} \frac{1}{u - m^{2}_{D^*} + im_{D^*}\Gamma_{D^*}} \left( -g^{\mu \nu} + \frac{(p_1 - p_4)^{\mu} (p_1 - p_4)^{\nu}}{m^{2}_{D^*}}  \right) (p_1 + p_4)_{\mu } (p_2 + p_3)_{\nu },  
\label{M6} 
\end{eqnarray}

	\begin{eqnarray}
 	\mathcal{M}_{(7a)} & = & \tau _{rs} ^{(i)} \tau _{r's'} ^{(j)} g^{2}_{\rho D^* D^*} \epsilon^{*\mu}_{3} \epsilon^{*\nu}_{4} \frac{1}{t - m^2_D} (2p_1 - p_3)_{\mu} (2p_2 - p_4)_{\nu},
 \nonumber \\
 	\mathcal{M}_{(7b)} & = & \tau _{rs} ^{(i)} \tau _{r's'} ^{(j)}  g^{2}_{\rho D^* D^*} \epsilon^{*\mu}_{4} \epsilon^{*\nu}_{3} \frac{1}{t - m^2_D} (2p_1 - p_4)_{\mu} (2p_2 - p_3)_{\nu}, \nonumber \\
\nonumber  \mathcal{M}^{ (7.c) }  & =  & g_{\rho D D^{*}}^{2} \varepsilon_{3}^{\alpha} \varepsilon_{4}^{\beta} \epsilon_{\mu \gamma \delta \alpha} \epsilon_{\nu \sigma \rho \beta}  \frac{1}{ t - m_{ D^{*} }^{2} + im_{ D^{*}} \Gamma_{ D^{*} } } 
	\left( - g^{\mu \nu} + \frac{ (p_1 – p_3 )^{\mu} (p_1 – p_3 )^{\nu} }{ m_{ D^{*} }^{2} } \right) p_{1}^{\delta} p_{3}^{\gamma} p_{2}^{\rho} p_{4}^{\sigma},
	\nonumber \\
 \mathcal{M}^{ (7.d) }  & = & - g_{\rho D D^{*}}^{2} \varepsilon_{3}^{\alpha} \varepsilon_{4}^{\beta} \epsilon_{\mu \gamma \delta \alpha} \epsilon_{\nu \sigma \rho \beta}  \frac{1}{ u - m_{ D^{*} }^{2} + im_{ D^{*}} \Gamma_{ D^{*} } } 
\left( - g^{\mu \nu} + \frac{ (p_1 – p_4 )^{\mu} (p_1 – p_4 )^{\nu} }{ m_{ D^{*} }^{2} } \right) p_{1}^{\gamma} p_{4}^{\delta} p_{2}^{\rho} p_{3}^{\sigma},
\label{M7} 
\end{eqnarray}
\end{widetext}
where $\tau _{rs} ^{(i,j)}$ is the isospin factor related to $i,j$-th component of $\pi,\rho$ isospin triplets and $r(s)$-th component of $D^{(\ast) }$ isospin doublets; $p_1$ and $p_2$ are the momenta of initial state particles, 
while $p_3$ and $p_4$ are those of final state particles; $s,t,u$ are the Mandelstam variables: $s = (p_1 +p_2)^2, t = (p_1 - p_3)^2,$ and $u = (p_1-p_4)^2$; and  $\epsilon_m  ^{(*)} \equiv \epsilon ^{(*)}(p_m) $ is the polarization vector.


\end{document}